\documentclass[pdflatex,sn-nature]{sn-jnl}

\usepackage{graphicx}%
\usepackage{multirow}%
\usepackage{amsmath,amssymb,amsfonts}%
\usepackage{amsthm}%
\usepackage{mathrsfs}%
\usepackage[title]{appendix}%
\usepackage{xcolor}%
\usepackage{textcomp}%
\usepackage{manyfoot}%
\usepackage{booktabs}%
\usepackage{algorithm}%
\usepackage{algorithmicx}%
\usepackage{algpseudocode}%
\usepackage{listings}%
\usepackage{lineno}

\usepackage{siunitx}
\DeclareSIUnit\corehour{\text{core-hours}}
\DeclareSIUnit\Bq{Bq}
\DeclareSIUnit\nsigma{\ensuremath{\sigma}}
\DeclareSIUnit[quantity-product = ]\percent{\char`\%}
\usepackage[nameinlink,capitalize]{cleveref}

\usepackage{upgreek}
\usepackage{nicematrix}
\usepackage{derivative} 
\usepackage{mathtools}
\usepackage{url}

\newcommand{\mymatht}[1]{\mbox{\ensuremath{#1}}}
\newcommand{\mymathtv}[1]{\ensuremath{#1}}

\newcommand{\myOrdernum}[1]{\ensuremath{\mathcal{O}(\num{#1})}}
\newcommand{\myOrderqty}[2]{\ensuremath{\mathcal{O}(\num{#1})\,\unit{#2}}}

\newcommand{\mytimes}{\ensuremath{\:\!\times\:\!}}
\newcommand{\myincrement}{\ensuremath{\Updelta}}
\newcommand{\myvect}[1]{\ensuremath{\boldsymbol{\mathrm{#1}}}}

\newcommand{\myintcramp}[2]{\smashoperator{\int\limits_{#1}^{#2}}}
\newcommand{\myiintcramp}[4]{\smashoperator{\int\limits_{#1}^{#2}}\smashoperator{\int\limits_{#3}^{#4}}}
\newcommand{\myintp}[2]{\int_{#1}^{#2}}

\newcommand{\mysum}[2]{\sum\limits_{#1}^{#2}}
\newcommand{\myprod}[2]{\prod\limits_{#1}^{#2}}

\newcommand{\myodd}[1]{\ensuremath{\mathrm{d}{#1}}}
\newcommand{\myoodd}[2]{\,\ensuremath{\mathrm{d}{#1}\,\mathrm{d}{#2}}}
\newcommand{\myodv}[3][]{\odv[#1]{#2}{#3}}
\newcommand{\myodvt}[2]{\mymatht{\mathrm{d}_{#2}{#1}}}
\newcommand{\mypdvv}[3]{\frac{\partial^2{#1}}{\partial{#2}\partial{#3}}}
\newcommand{\mypdvvt}[3]{\mymatht{\partial_{#2}\partial_{#3}{#1}}}

\newcommand{\myNat}{\mathbb{N}}
\newcommand{\myReal}{\mathbb{R}}
\newcommand{\myReall}[1]{\mathbb{R}_{#1}}
\newcommand{\myNatu}[1]{\mathbb{N}^{#1}}
\newcommand{\myRealu}[1]{\mathbb{R}^{#1}}
\newcommand{\myRealul}[2]{\mathbb{R}^{#1}_{#2}}

\newcommand{\mycode}[1]{\texttt{#1}}

\newcommand{\myversion}[1]{{#1}}
\newcommand{\mycampaign}[1]{\texttt{#1}}

\newcommand{\mygammaf}[1]{\ensuremath{\operatorname{\Upgamma}\!\left(#1\right)}}
\newcommand{\mydomain}[1]{\ensuremath{\mathcal{D}{#1}}}
\newcommand{\myrandomvec}[1]{\ensuremath{\boldsymbol{\mathit{#1}}}}

\newcommand{\mynuclt}[3]{\ensuremath{\cramped{\prescript{#1}{#2}{\mathrm{#3}}}}}
\newcommand{\mynucltbf}[3]{\ensuremath{\boldsymbol{\cramped{\prescript{#1}{#2}{\mathrm{#3}}}}}}
\newcommand{\myBa}{\mynuclt{133}{56}{Ba}}
\newcommand{\myCs}{\mynuclt{137}{55}{Cs}}

\newcommand{\myBabf}{\mynucltbf{133}{56}{Ba}}
\newcommand{\myCsbf}{\mynucltbf{137}{55}{Cs}}

\newcommand{\myUnat}{\ensuremath{\cramped{\mathrm{U}_{\mathrm{nat}}}}}
\newcommand{\myThnat}{\ensuremath{\cramped{\mathrm{Th}_{\mathrm{nat}}}}}
\newcommand{\myKnat}{\ensuremath{\cramped{\mathrm{K}_{\mathrm{nat}}}}}
\newcommand{\myRnIsotope}{\mynuclt{222}{86}{Rn}}

\newcommand{\mydispersion}{\ensuremath{\alpha_{\scriptscriptstyle\text{NB}}}}

\newcommand{\myCsstr}{\ensuremath{\xi_{\scriptscriptstyle\text{Cs-137}}}}
\newcommand{\myBastr}{\ensuremath{\xi_{\scriptscriptstyle\text{Ba-133}}}}
\newcommand{\myKnatstr}{\ensuremath{\xi_{\scriptscriptstyle\text{K-nat}}}}
\newcommand{\myUnatstr}{\ensuremath{\xi_{\scriptscriptstyle\text{U-nat}}}}
\newcommand{\myThnatstr}{\ensuremath{\xi_{\scriptscriptstyle\text{Th-nat}}}}
\newcommand{\mydRnstr}{\ensuremath{\myincrement{\xi_{\scriptscriptstyle\text{Rn-nat}}}}}

\newcommand{\myonlinecite}[1]{Ref.~\cite{#1}}
\newcommand{\myonlinecitepl}[1]{Refs.~\cite{#1}}

\begin{document}

\title{%
\begin{center}
Quantitative mobile gamma-ray spectrometry\\
through Bayesian inference
\end{center}
}

\author*[1,2,]{\fnm{David} \sur{Breitenmoser}}\email{david.breitenmoser@psi.ch}
\author[1]{\fnm{Alberto} \sur{Stabilini}}
\author[1]{\fnm{Malgorzata Magdalena} \sur{Kasprzak}}
\author[1]{\fnm{Sabine} \sur{Mayer}}
\affil[1]{\orgdiv{Department of Radiation Safety and Security}, \orgname{Paul~Scherrer~Institute (PSI)}, \orgaddress{\street{Forschungsstrasse~111}, \city{Villigen~PSI}, \postcode{AG 5232}, \country{Switzerland}}}
\affil[2]{\orgdiv{Department of Nuclear Engineering \& Radiological Sciences}, \orgname{University of Michigan}, \orgaddress{\street{2355 Bonisteel Blvd.}, \city{Ann Arbor}, \postcode{MI 48109-2104}, \country{United States of America}}}


\abstract{\mathversion{normal}%
Accurate quantitative mapping of gamma-ray emitters is critical for applications ranging from radiological emergency response and environmental monitoring to nuclear security and deep space exploration. Here, we show that such mapping can be achieved by combining mobile gamma-ray spectrometry with high-fidelity Monte Carlo simulations and full-spectrum Bayesian inference. Using \qty{6}{\s} of single-pass mobile spectrometry data benchmarked against independent \textit{in-situ} and laboratory assays, we demonstrate decisive source mixture identification (\mbox{$>\!\!5\sigma$}), meter-scale source localization, and recovery of source activities with percent-level accuracy. The developed method marks a critical advance in quantitative gamma-ray sensing, enabling improved radiological situational awareness, enhanced terrestrial geophysical and geochemical mapping, as well as more robust constraints on radionuclide abundances on extraterrestrial bodies across the Solar System.
}

\keywords{Bayesian inference, Monte Carlo simulation, gamma-ray spectroscopy, inorganic scintillator, Markov chain Monte Carlo}

\maketitle

\section{Introduction}\label{sec:Introduction}

Mobile gamma-ray spectrometry (MGRS) has emerged as a powerful technique for detecting and characterizing gamma-ray sources in diverse environments. By deploying spectrometers on ground-based, airborne, marine, and spaceborne platforms, MGRS enables rapid inference of source mixtures, source localizations, and associated emission rates in a wide range of fields, including environmental monitoring of marine ecosystems \cite{Jones2001a,Lee2023c}, mapping radioactive contamination after nuclear accidents \cite{Rosenthal1991a,Drovnikov1997a,Sanada2014a}, nuclear security and nonproliferation \cite{Deal1972a,Hellfeld2021a,Curtis2020a,Salathe2021a}, geophysics \cite{Fishman1994,Tavani2011,Smith2011a,Neubert2020a,Sinclair2011a,Baldoncini2017a}, and planetary science \cite{Prettyman2006a,Hahn2007a,Kobayashi2010a,Peplowski2011b,Peplowski2016c,Prettyman2017,Prettyman2019a}. More recently, the advent of uncrewed mobile platforms has opened new opportunities to deploy MGRS systems in previously inaccessible or hazardous environments, further broadening its scope of application \cite{Sanada2015a,Connor2016a,Naumenko2018,Chen2020a,Sinclair2021}.

While MGRS has shown considerable promise for localizing gamma-ray sources, robust identification of source mixtures and accurate quantification of their activities remain major challenges. In contrast to \textit{in-situ} and laboratory settings, MGRS systems operate in a near real-time data acquisition mode with sampling frequencies as low as \myOrderqty{1}{\Hz} and source--detector distances ranging from \myOrderqty{1d2}{\m} in terrestrial surveys to at least \myOrderqty{1d2}{\kilo\m} for spaceborne platforms. Consequently, identifying an unknown source mixture and inferring the corresponding source activities from MGRS data constitutes a severely ill-posed inverse problem. The combination of low-count statistics, variations in source--detector separation and relative orientation induced by platform motion, as well as strong spectral correlations between different radionuclides leads to non-uniqueness and instability in the solution space. To address this ill-posed inverse problem, the sparse and low-count spectra characteristic of most MGRS systems preclude the use of conventional peak-fitting methods, which are only effective in stationary or otherwise well-controlled source--detector settings, such as prolonged planetary surveys \cite{Hahn2007a,Kobayashi2010a,Prettyman2017,Prettyman2019a}. Instead, template-matching approaches, commonly referred to as full-spectrum analysis (FSA), have emerged as the most viable alternative and have demonstrated promising results in both terrestrial \cite{Grasty1985a,Minty1998d,Hendriks2001a} and spaceborne applications \cite{Prettyman2006a,Prettyman2011a,Peplowski2016c}. However, current FSA implementations remain limited by systematic biases in the inferred source strengths and reduced sensitivity under the low-count conditions characteristic for single-pass MGRS surveys \cite{Peplowski2016c,Ohera2024,Sinclair2016a}.

We attribute these limitations of current FSA implementations to two primary factors: inaccuracies in spectral-template generation and limitations of the statistical models employed in the inversion. On the template side, spectral-template libraries---defined here as collections of reference detector response spectra for specific radionuclide sources and source--detector configurations---may be derived either from empirical calibration measurements or from numerical simulations. Empirically derived libraries are necessarily sparse, being restricted to a limited set of radionuclide sources and geometries, and their extension to broader survey conditions is prohibitively costly and time-consuming, often requiring days of dedicated measurements and extensive source handling \citep{Grasty1991a,Minty1990a}. Moreover, such experimentally acquired templates are unavoidably affected by variable background contributions, introducing additional systematic uncertainties. Numerically generated templates, by contrast, offer substantially greater flexibility but have historically relied on oversimplified Monte Carlo models in which the mobile platforms were either neglected entirely or represented by highly simplified geometries to reduce computational complexity \citep{Allyson1998a,Sinclair2011a,Torii2013a,Sinclair2016a,Kulisek2018a,Curtis2020a}. These approximations fail to reproduce key platform-dependent effects such as scattering and attenuation within the platform structure, resulting in persistent, energy-dependent template biases that can exceed \qty{200}{\percent} at photon energies below \qty{\sim100}{\keV} \citep{Allyson1998a,Kulisek2018a,Curtis2020a}. Recent advances in high-fidelity Monte Carlo modeling have demonstrated that incorporating detailed dynamic mass models into the simulations can substantially reduce template-related discrepancies, in particular at low photon energies \qty{\leq200}{\keV} \cite{Breitenmoser2025a,Breitenmoser2026}.

The second limitation arises from the inversion framework itself. Most existing FSA pipelines adopt frequentist formulations based on maximum likelihood estimation (MLE) algorithms combined with Gaussian likelihoods \cite{Minty1992a,Grasty1985a,Minty1998d,Hendriks2001a,Sinclair2011a,Prettyman2006a}. While computationally efficient, Gaussian likelihoods are not ideally suited to the sparse counting statistics characteristic of MGRS measurements and may lead to biased parameter estimates and unreliable uncertainty quantification under low-count conditions \cite{Humphrey2009,Yamada2019}. To better represent the discrete nature of radiation counting data, Poisson-based MLE formulations have subsequently been proposed \cite{Salathe2021a}. However, beyond the choice of likelihood model, the inference of source mixtures, activities, and locations from sparse MGRS measurements remains a challenging high-dimensional and ill-posed inverse problem. In other scientific disciplines involving similarly complex inverse problems, including gravitational-wave astronomy \cite{Veitch2015,Ashton2019,Smith2020} and exoplanet research \cite{MacDonald2017,Pinhas2018}, Bayesian inference has emerged as a powerful alternative to traditional frequentist approaches by providing a unified probabilistic framework for combining prior knowledge with measurement data. In particular, Bayesian methods enable rigorous uncertainty quantification, explicit characterization of parameter correlations, and the natural incorporation of nuisance parameters into the inference process \cite{Trotta2008a,VonToussaint2011a}. In MGRS, where multiple correlated source parameters must be inferred from sparse spectral observations under dynamically varying conditions, these capabilities provide a strong motivation for adopting a Bayesian framework.

In this work, we realize such a Bayesian framework for quantitative MGRS by integrating probabilistic inversion with high-fidelity numerical template generation. The numerical templates leverage detailed, dynamically updated Monte Carlo models of the measurement platform to accurately predict spectral responses across a wide range of source--detector configurations. Within this framework, Bayesian inference enables the joint estimation of source mixtures, associated activities, and source locations for an arbitrary number of radionuclides while providing rigorous uncertainty quantification and characterization of parameter correlations. We evaluate the proposed methodology on a series of dedicated MGRS measurements benchmarked against laboratory and \textit{in-situ} reference data, under both static and dynamic source--detector configurations, demonstrating accurate and reliable inference of natural and anthropogenic radionuclides. By integrating high-fidelity forward modeling with Bayesian inference, this framework provides a generalizable approach for quantitative analysis of sparse and dynamically acquired MGRS data, with applications spanning radiological emergency response, planetary exploration, nuclear security, and geophysics.

\section{Results}\label{sec:Results}

\subsection{Full-spectrum Bayesian inference framework}\label{subsec:fsbi}

\begin{figure}[ht]
\centering
\includegraphics[width=0.9\textwidth]
{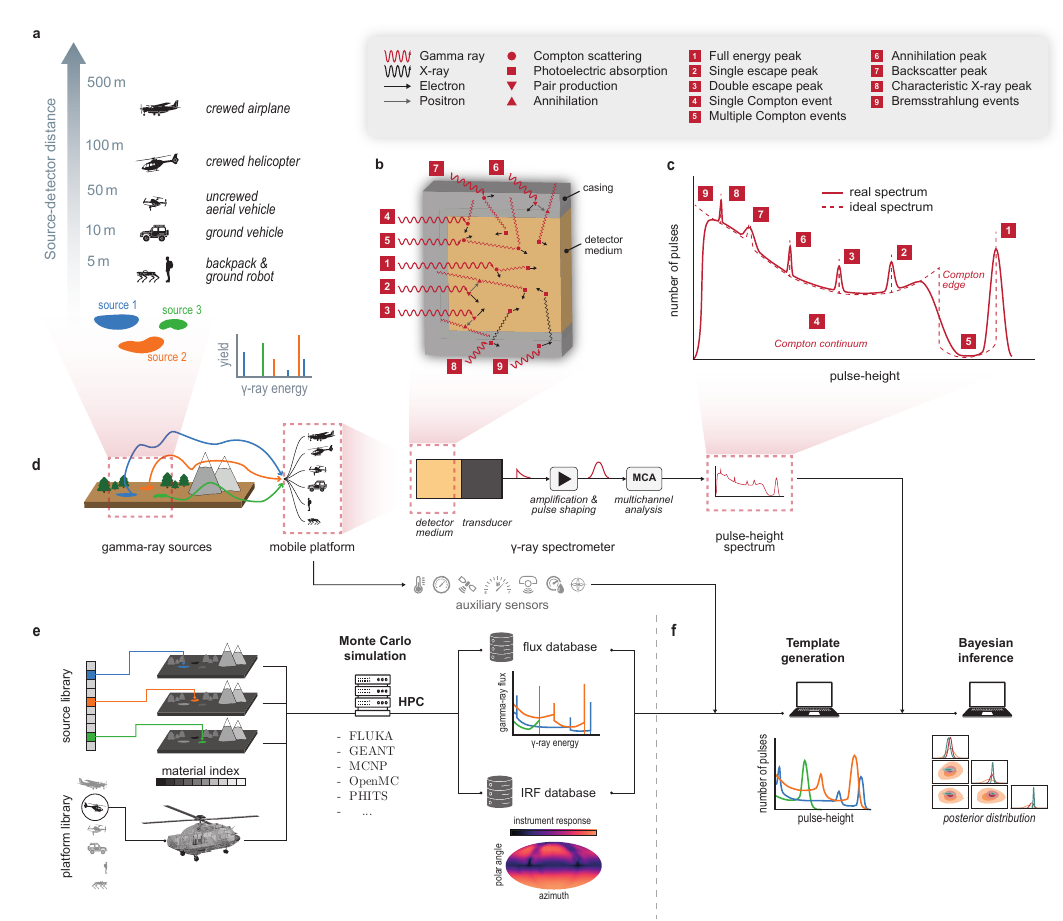}
\caption{\textbf{Overview of the full-spectrum Bayesian inference framework.} \textbf{a}~Characteristic source--detector distances for various mobile platforms utilized in MGRS. \textbf{b}~Gamma-ray-matter interaction mechanisms and associated secondary processes. \textbf{c}~Characteristic spectral features in the full-spectrum pulse-height response to monoenergetic gamma-ray events. \textbf{d}~Hierarchical  sequence of physical processes involved in MGRS. \textbf{e}~Instrument response function (IRF) and double differential gamma-ray flux database generation utilizing high-fidelity radiation transport codes \cite{Romano2015,Allison2016,Goorley2016,Ahdida2022a,Sato2024} run on high-performance computing (HPC) infrastructure. \textbf{f}~High-fidelity spectral template generation and Bayesian inference on a local workstation utilizing general-purpose programming languages and dedicated Bayesian numerical codes \cite{Marelli2014a,Carpenter2017,Buchner2021c,Abril-Pla2023}.}
\label{fig:FrameworkOverview}
\end{figure}

As illustrated in \cref{fig:FrameworkOverview}, our proposed FSA methodology combines MGRS pulse-height spectral data with high-performance computing (HPC) based template generation in a unified Bayesian framework. Bayesian inference offers a natural, consistent, and transparent way of combining existing information with empirical data to solve complex inverse problems using a solid probabilistic decision theory framework \cite{Trotta2008a,Gelman2013,VonToussaint2011a,DAgostini2003a}. Within this Bayesian FSA formalism, the inverse problem consists of jointly inferring the gamma-ray source strengths \mymatht{\Xi=\{\xi_s\in\myReall{+}\}^{N_s}_{s=1}} and source locations \mymatht{\mathcal{X}=\{\myvect{x}_s\in\myRealu{3}\}^{N_s}_{s=1}} for an assumed source mixture \mymatht{\mathcal{S}}, comprising \mymatht{N_s\in\myNat} sources whose identities are specified \textit{a priori}. The inference is performed using a set of \mymatht{N_\mathcal{Y}\in\myNat} observed gamma-ray spectra \mymathtv{\mathcal{Y}=\{\myvect{y}_i\in\myNatu{N_y}\}^{N_\mathcal{Y}}_{i=1}}, where each spectrum \mymatht{\myvect{y}_i} contains measured counts in \mymatht{N_y\in\myNat} spectral channels. The corresponding acquisition is carried out under known experimental conditions \mymatht{\myvect{\mathfrak{d}}_t \in \myRealu{N_\mathfrak{d}}}, which encode the known platform and environmental state parameters, e.g., the platform position and atmospheric pressure, with the full set of \mymatht{N_\mathfrak{D}\in\myNat} experimental conditions denoted as \mymatht{\mathfrak{D}=\{\myvect{\mathfrak{d}}_t \in \myRealu{N_\mathfrak{d}}\}^{N_\mathfrak{D}}_{t=1}}. Probabilistic inference of the unknown model parameters is achieved by applying Bayes’ theorem

\begin{equation}
    \pi\left(\myvect{\uptheta}\mid \mathcal{Y},\mathfrak{D},\mathcal{S}\right) = \frac{\mathcal{L}\left(\myvect{\uptheta};\mathcal{Y},\mathfrak{D},\mathcal{S}\right)\pi\left(\myvect{\uptheta}\mid\mathcal{S}\right)}{\myintp{\mydomain{_{\myrandomvec{\Theta}}}}{}
    \mathcal{L}\left(\myvect{\uptheta};\mathcal{Y},\mathfrak{D},\mathcal{S}\right)\pi\left(\myvect{\uptheta}\mid\mathcal{S}\right)\myodd{\myvect{\uptheta}}} ,
\label{eq:BayesTheorem}
\end{equation}

\noindent where \mymatht{\mathcal{L}(\myvect{\uptheta};\mathcal{Y},\mathfrak{D},\mathcal{S})} and \mymatht{\pi(\myvect{\uptheta}\mid\mathcal{S})} denote the likelihood and parameter prior defined on the \mymatht{M}-dimensional parameter space \mymatht{\Uptheta=\{\boldsymbol{\uptheta}\in\myRealu{M}\}} \cite{Trotta2008a}. The full parameter vector \mymatht{\boldsymbol{\uptheta}} is formed by stacking the source strengths \mymatht{\Xi}, source locations \mymatht{\mathcal{X}}, and additional nuisance parameters into a single joint parameter vector, i.e., \mymatht{\boldsymbol{\uptheta} \equiv (\Xi,\mathcal{X},\eta)}, where \mymatht{\eta} collects the set of additional nuisance parameters required by the forward model. This construction enables simultaneous inference and consistent uncertainty propagation across all model components.

The resulting posterior distribution \mymatht{\pi(\myvect{\uptheta}\mid \mathcal{Y},\mathfrak{D},\mathcal{S})} provides a complete probabilistic description of the parameter space conditioned on the observed data, experimental conditions, and assumed source-mixture model. Two critical components of this framework are the likelihood function and prior specifications, which together must faithfully reflect the statistical character of MGRS count data and existing domain knowledge. Drawing on treatments in cosmology, nuclear, and particle physics \cite{Breitenmoser2025j,Praszalowicz2011,Tezlaf2023,Perez2021,Fry2013,Hurtado-Gil2017,Hameeda2021}, we adopt a negative binomial likelihood to generalize the stationary Poisson assumption in prior FSA formulations to the nonstationary count processes characteristic of MGRS, complemented by weakly informative priors to avoid overly restrictive assumptions on the model parameters. The complete likelihood formulation and prior specifications are detailed in \cref{subsec:bayestheorymethod}.

To determine which source mixture \mymatht{\mathcal{S}} is most consistent with the data \mymatht{\mathcal{Y}}, we perform Bayesian model comparison across a finite collection of \mymatht{N_\mathcal{S}\in\myNat} competing source-mixture models \mymatht{\{\mathcal{S}_k\}_{k=1}^{N_\mathcal{S}}} with associated model priors \mymatht{\pi(\mathcal{S}_k)}. Assuming noninformative model priors, i.e., \mymathtv{p(\mathcal{S}_k) = 1/N_\mathcal{S} \;  \forall k}, the relative support for one model over another is quantified by the Bayes factor \cite{Trotta2008a,VonToussaint2011a}

\begin{equation}
\mathcal{B}_{ij} = \frac{\pi( \mathcal{Y} \mid \mathcal{S}_i,\mathfrak{D})}{\pi(\mathcal{Y} \mid \mathcal{S}_j,\mathfrak{D})} ,
\label{eq:BayesFactor}
\end{equation}

\noindent where \mymatht{\mathcal{Z}_i\coloneqq\pi(\mathcal{Y}\mid\mathcal{S}_i,\mathfrak{D})} denotes the Bayesian evidence of model \mymatht{\mathcal{S}_i}, corresponding to the denominator term in \cref{eq:BayesTheorem} \cite{Trotta2008a}.

While both the posterior and Bayesian evidence are analytically intractable for general likelihood and prior distributions \cite{Gelman2013}, advances in Bayesian computation over the past decades have enabled their numerical estimation with controllable accuracy and precision. State-of-the-art approaches, including Markov chain Monte Carlo (MCMC) methods \cite{Metropolis1953,Hastings1970,Foreman-Mackey2013a,Goodman2010} combined with MCMC-based evidence estimators \cite{Perrakis2014,Metodiev2024,Llorente2023}, as well as nested sampling \cite{Skilling2006a,Feroz2009a,Speagle2020a,Buchner2021c,Ashton2022a}, can now be employed to efficiently explore posterior distributions and estimate Bayesian evidences, even for highly complex inverse problems. However, these methods remain computationally demanding, often requiring \mymatht{\myOrdernum{1d4}} likelihood evaluations for a single inference task in MGRS-like inverse problems, depending on convergence and precision requirements (see \cref{subsec:bayescompmethod}). Consequently, the development of likelihood models that can be evaluated rapidly is a key prerequisite for the practical implementation of the proposed Bayesian FSA methodology.

Constructing a likelihood model \mymatht{\mathcal{L}} that can be evaluated rapidly while retaining predictive accuracy for the full-spectrum MGRS response under varying experimental conditions is a computationally challenging task. As illustrated in \cref{fig:FrameworkOverview}, the spectral response of an MGRS system arises from a hierarchical sequence of physical processes: the emission characteristics of multiple gamma-ray sources within the field of view, spanning distances from \myOrderqty{1d1}{\m} to \myOrderqty{1d2}{\kilo\m}; the interactions and propagation of the emitted gamma rays through the environment and the mobile platform, including the generation of secondary X-rays and annihilation photons; and the energy-conversion chain in the detector system, spanning processes across multiple spatial and temporal scales, including microscopic energy deposition, charge-carrier thermalization, scintillation-light production, and subsequent signal formation and processing \cite{VasilEv2014a,Breitenmoser2023c}. As discussed in \cref{sec:Introduction}, Monte Carlo simulations \cite{Allyson1998a,Sinclair2011a,Torii2013a,Sinclair2016a,Kulisek2018a,Curtis2020a,Salathe2021a} utilizing high-fidelity radiation transport codes \cite{Romano2015,Allison2016,Goorley2016,Ahdida2022a,Sato2024} represent the most established and reliable method for numerically approximating the spectral response of MGRS systems. Their main drawback, however, is computational cost: high-fidelity simulations typically require \ensuremath{\Updelta{t}_{\mathrm{MC}}=\myOrderqty{1d4}{\corehour}} per evaluation in MGRS \cite{Breitenmoser2025a}. Since template generation scales with both the number of sources \mymatht{N_s} and the number of unique experimental states \mymatht{N_\mathfrak{D}}, brute-force Monte Carlo approaches are prohibitive even for moderate survey sizes, despite access to current HPC infrastructure. 

To overcome this limitation, we employ a two-stage forward-modeling strategy: (i) precomputing instrument response functions (IRFs) and gamma-ray flux libraries using HPC infrastructure for a set of predefined experimental conditions \mymatht{\bigcup_{t=1}^{N_\mathfrak{D}} \myvect{\mathfrak{d}}_t} and source-component sets \mymatht{\bigcup_{k=1}^{N_\mathcal{S}} \mathcal{S}_k}, and (ii) assembling spectral templates from these precomputed components on a local workstation (see \cref{subsec:templatemethod}). As demonstrated in our previous work \cite{Breitenmoser2026}, this approach reduces the evaluation cost to \ensuremath{\myOrderqty{1d0}{\s}}, enabling high-fidelity likelihood evaluations at practical computational expense and, in turn, providing a viable pathway to perform full-spectrum Bayesian inference in MGRS.

\subsection{Data selection}\label{subsec:dataresult}

We evaluated the proposed Bayesian methodology on the Swiss Airborne Gamma-Ray Spectrometry (SAGRS) system, a state-of-the-art airborne MGRS platform equipped with a \qty{\sim1.7d4}{\cubic\cm} NaI(Tl) spectrometer mounted in an A\'{e}rospatiale AS332M1 Super Puma helicopter (see~\cref{subsec:sagrs}). The system offers both a validated high-fidelity Monte Carlo mass model and a curated archive of systematically collected datasets spanning more than five years in Switzerland and abroad \cite{Breitenmoser2022,Breitenmoser2025a}, providing a robust basis for benchmarking and validation.

Here, we focus on two benchmark configurations acquired during the \mycampaign{ARM22} exercise at the Thun military training ground (\qty{46.753}{\degree}N, \qty{7.596}{\degree}E) in Switzerland on June 16, 2022 \cite{Butterweck2023}. Both configurations employed the same two sealed anthropogenic radionuclide point sources ({\myCs} and {\myBa}) with bin-mode acquisition at \qty{1}{\Hz}. In the first configuration, the SAGRS system operated in a static hover-flight mode for \qty{\sim5}{\minute}, with the helicopter positioned at a ground clearance of \qty{\sim90}{\m}, such that the individually deployed radionuclide point sources on the ground were aligned with respect to the aircraft vertical symmetry axis. In the second configuration, the SAGRS system performed single-pass surveys at a ground clearance of \qty{\sim30}{\m} and a flight speed of \qty{\sim10}{\m\per\s}, with flight tracks passing directly over the ground-based source positions, for three source scenarios consisting of individually deployed {\myCs} and {\myBa} point sources, as well as a combined dual-source deployment (co-located at the same position with a spatial separation \mymatht{<\!\qty{20}{\cm}}).

The exercise was specifically designed for reproducibility and quantitative validation. Reference activities of the anthropogenic sources were determined from calibration certificates based on laboratory gamma-ray assays. Natural background activity concentrations of terrestrial potassium {\myKnat}, uranium {\myUnat} (with progeny), and thorium {\myThnat} (with progeny) were constrained by \textit{in-situ} gamma-ray spectroscopic measurements at \num{23} reference locations across the training ground \cite{Butterweck2021a}.

Data reduction to retrieve pulse-height count spectra and associated calibration models (energy and spectral resolution) was performed independently for each experiment using the post-processing pipelines \mycode{RLLSpec} and \mycode{RLLCal} \cite{Breitenmoser2025a}. The four independently read-out scintillation crystals were treated jointly as a single detector response in the present analysis, yielding spectra with \num{1008} channels spanning an energy range from \qty{\sim50}{\keV} to \qty{\sim3}{\MeV} (see~\cref{subsec:sagrs}). Helicopter states were retrieved from onboard auxiliary sensors. Meteorological parameters were taken from the MeteoSwiss automatic weather station on the Thun training ground (WIGOS-ID: 0-20000-0-06731). Each dataset comprises a set of gross-count spectra \mymatht{\mathcal{Y}} (\mymatht{N_\mathcal{Y}=1} for the hover-flight benchmark and \mymatht{N_\mathcal{Y}=6} for the single-pass benchmark) with associated measurement live times \mymatht{\myincrement{t}_\mathrm{gr}} and a set of known experimental conditions \mymatht{\mathfrak{D}} that encodes the platform and environmental states relevant to the likelihood evaluation: helicopter position and orientation (pose), fuel level, and ambient conditions (air temperature, pressure, humidity). A listing of retrieved experimental-condition values is provided in Supplementary Table~1. Further details on the measurement setup, acquisition, and data processing are available in \myonlinecite{Breitenmoser2025a}.

\subsection{Inferring natural and anthropogenic radionuclide activities under controlled hover-flight conditions}\label{subsec:hfresults}

We first evaluate the proposed Bayesian inversion framework under the controlled hover-flight configuration, restricting the inference to radionuclide activities while fixing the source locations to their known experimental positions. The hover configuration provides a well-constrained benchmark and is particularly well suited for validation, as the stationary source alignment enables a robust assessment of the forward model and likelihood formulation under minimal geometric uncertainty. 

We solve the inverse problems for each of the two deployed anthropogenic source scenarios separately, using the Bayesian inversion framework outlined in \cref{subsec:fsbi}. Specifically, we sample the posterior distribution via Bayes' theorem in \cref{eq:BayesTheorem} using an affine-invariant ensemble MCMC algorithm \cite{Goodman2010} (see \cref{subsec:bayescompmethod}) to jointly infer the activities of the individual anthropogenic point sources as well as the source activity concentrations of the extended naturally occurring terrestrial radionuclides ({\myKnat}, {\myThnat}, and {\myUnat}) with their progeny. Combined spectral contributions of residual intrinsic, cosmic-ray, and atmospheric radon progeny backgrounds were estimated empirically from independent background flights over Lake Thun at orthometric heights matched to the hover measurements (see Supplementary Fig.~1). Because these flights were conducted within \mymatht{\qty{10}{\km}} and less than \qty{1}{\hour} of the experiments, we considered them representative of the intrinsic and cosmic background at the training site. To account for spatiotemporal variability in radon progeny concentrations, we included an additional variable radon source term \mymatht{\mydRnstr\in\myReal} and inferred it jointly with the other source strengths in the Bayesian analysis (see \cref{subsec:fluxmethod}).

\begin{figure}[tb!]
\centering
\includegraphics[width=1.0\textwidth]{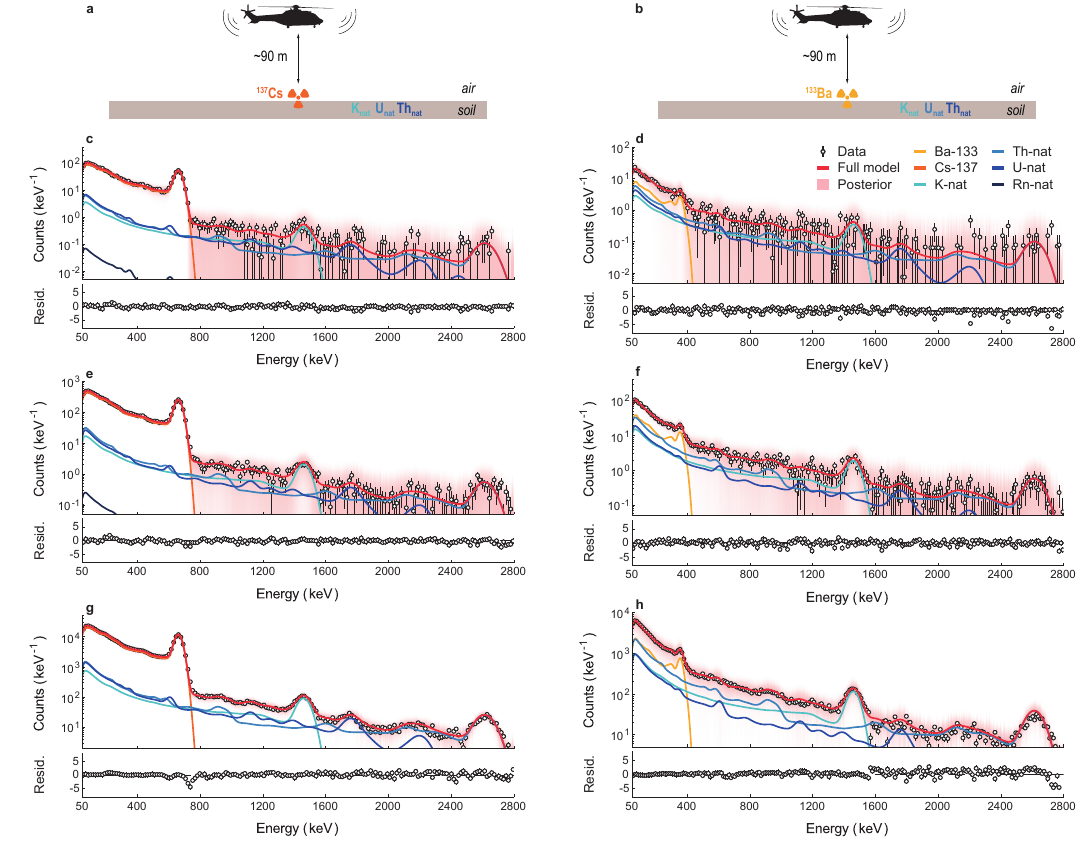}
\caption{\textbf{Posterior predictive distribution results.} Posterior predictive distributions are shown alongside the measured pulse-height spectra for two measurement configurations: \textbf{a}~Hover flight above an anthropogenic {\myCs} source. \textbf{b}~Hover flight above an anthropogenic {\myBa} source. Results are displayed for three live times: \qty{1}{\s} (\textbf{c}, \textbf{d}), \qty{5}{\s} (\textbf{e}, \textbf{f}), and the full acquisition \qty{\sim5}{\minute} (\textbf{g}, \textbf{h}). Uncertainties in the measured pulse-height spectra are shown as 1~standard errors. Residuals are shown as standardized posterior predictive residuals. Posterior predictive distributions are accompanied by point predictions based on the posterior median estimates. Spectral signatures scaled by the posterior median source strengths and measurement live time are also indicated for all sources included in the forward model, namely the anthropogenic {\myCs} and {\myBa} sources; the three natural terrestrial radionuclides {\myKnat}, {\myThnat}, and {\myUnat}; and the absolute radon source term (partially visible in \textbf{c} and \textbf{e}). For improved interpretability, all spectral quantities were corrected for residual backgrounds, and measured spectra were additionally rebinned from the native \qty{\sim3}{\keV} channel spacing  used for the Bayesian inference to an effective bin width of \qty{\sim12}{\keV} (for {\myCs}) and \qty{\sim15}{\keV} (for {\myBa}). All spectra shown in energy units are derived from calibrated pulse-height channels for visualization only. The line drawings of the AS332M1 helicopter in panels \textbf{a--b} were adapted from Jetijones, \href{https://creativecommons.org/licenses/by/3.0}{\texttt{CC~BY~3.0}}, via Wikimedia Commons.}\label{fig:PostPred}
\end{figure}

\begin{figure}[tbh!]
\centering
\includegraphics[width=0.98\textwidth]{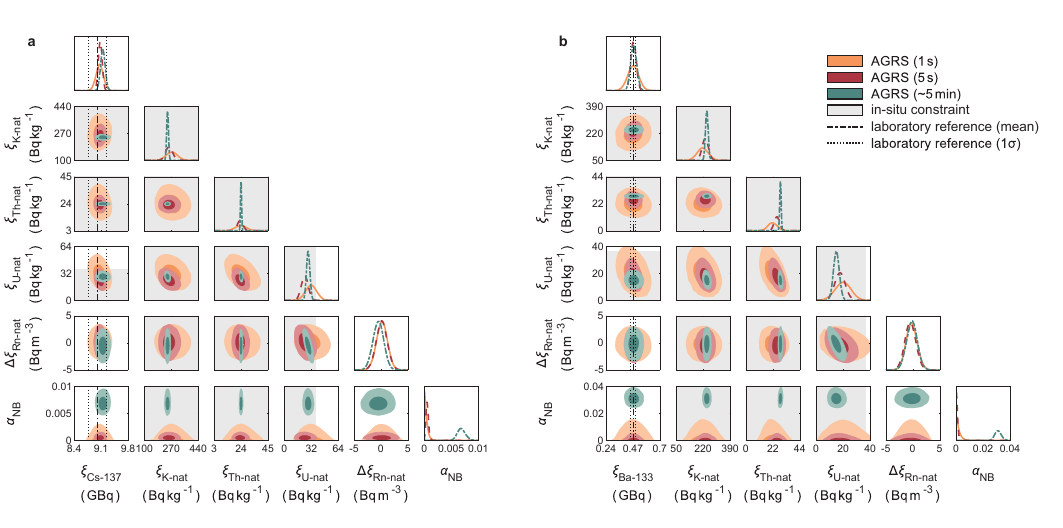}
\caption{\textbf{Posterior distribution results.} Posterior distributions are shown for two measurement configurations: \textbf{a}~Hover flight above an anthropogenic {\myCs} source. \textbf{b}~Hover flight above an anthropogenic {\myBa} source. Results are displayed for three live times: \qty{1}{\s}, \qty{5}{\s}, and \qty{\sim5}{\minute}. Each subfigure shows posterior distributions for the related model parameters: source activities of the anthropogenic {\myCs} and {\myBa} point sources ({\myCsstr}, {\myBastr}); activity mass concentrations of the three natural terrestrial radionuclides {\myKnat}, {\myThnat}, and {\myUnat} ({\myKnatstr}, {\myThnatstr}, {\myUnatstr}); activity volume concentration of the radon source term ({\mydRnstr}), as well as the dispersion parameter of the negative binomial distribution ({\mydispersion}). Off-diagonal panels depict bivariate posterior marginals with \qty{68}{\percent} and \qty{99}{\percent} probability contours, while diagonal panels show the corresponding univariate marginals. Gray-shaded regions indicate \textit{in-situ} gamma-ray spectroscopy constraints on terrestrial radionuclide activity concentrations, and black dashed lines mark best-estimate laboratory gamma-ray assay activities for the two anthropogenic sources with \mymatht{1\upsigma} uncertainty bounds.}
\label{fig:CornerPlot}
\end{figure}

We present the results from these Bayesian computations as posterior predictive distributions in \cref{fig:PostPred} alongside univariate and bivariate marginal posterior distributions shown in \cref{fig:CornerPlot} for acquisition times of \qty{1}{\s}, \qty{5}{\s}, and the full measurement duration (\qty{\sim5}{\minute}), spanning the range commonly encountered in MGRS applications \cite{Lyons2012,Sanada2014a,Winkelmann2004a,Peplowski2016c}. We assess the accuracy of the inferred anthropogenic source activities for {\myCs} and {\myBa} by comparing the posterior median activities with independently determined reference activities (see \cref{subsec:dataresult}). As shown in \cref{fig:CornerPlot}, the posterior predictions are in excellent agreement with the reference values, with relative deviations below \mymatht{\qty{2}{\percent}}. Importantly, all deviations remain statistically insignificant within the combined \mymatht{1\sigma} uncertainties of the posterior estimates and the reference activities.

Assessing the accuracy of the inferred activities for the terrestrial primordial radionuclides {\myKnat}, {\myThnat}, and {\myUnat} is inherently more challenging, owing to the substantially larger statistical and systematic uncertainties associated with both the posterior estimates and the reference activities obtained from \textit{in-situ} gamma-ray spectroscopy. These uncertainties arise primarily from spatial heterogeneity in the natural radionuclide distribution and limited counting statistics at higher spectral energies. Nevertheless, for all experiments, the posterior activity estimates are statistically consistent with the corresponding \textit{in-situ} constraints as well as with each other within their combined uncertainty ranges (see Supplementary Tables~3 and 4). This consistency confirms the robustness of the proposed Bayesian framework for quantifying natural background radionuclides under field conditions, despite the reduced accuracy that is fundamentally imposed by the available reference information.

The radon source term \mymatht{\mydRnstr} and the dispersion parameter \mymatht{\mydispersion} governing overdispersion in the likelihood function (see \cref{subsec:bayestheorymethod}) cannot be measured independently and therefore have no external reference values. Both quantities are consequently treated as latent parameters and inferred directly from the observational data within the Bayesian framework. In the case of radon, the inferred source term remains statistically consistent with the empirically estimated atmospheric background, indicating stable radon progeny concentrations between the background flights and the hover measurements. In contrast, the dispersion parameter \mymatht{\mydispersion} exhibits a systematic dependence on acquisition duration, with larger values for the full-length datasets compared to short integration times, corresponding to a statistically significant separation between both regimes at the \mymatht{>\!\!8\sigma} level. This behavior indicates increasing sensitivity to departures from the ideal Poisson counting model as the statistical uncertainty decreases with longer acquisition times. At short integration times, statistical fluctuations may dominate over small model discrepancies, whereas longer acquisitions provide sufficient counting statistics to reveal residual inconsistencies between the observations and the assumed forward model. While the forward model accounts for dominant second-order environmental effects, including atmospheric contributions and ground back-scattering (see \cref{subsec:fluxmethod}), residual effects such as higher-order scattering contributions, detector response deviations, or other unmodeled measurement variations may contribute to the observed trend.

\subsection{Inferring source mixtures, locations, and activities in single-pass surveys}\label{subsec:spresults}

In the previous section, we demonstrated accurate activity quantification under known source locations and source mixtures, using a single integrated spectrum acquired under a stationary source–detector configuration. Here, we extend the inference task to the more demanding and operationally realistic single-pass survey configuration, in which source mixture, source location, and source activities are retrieved simultaneously from dynamic flight data acquired at a ground clearance of \qty{\sim30}{\m} and a flight speed of \qty{\sim10}{\m\per\s}, across three source scenarios: a single {\myBa} source, a single {\myCs} source, and a combined dual-source {\myBa}-{\myCs} deployment (co-located at the same position with a spatial separation \mymatht{<\!\qty{20}{\cm}}). In all cases, the inferred source mixtures additionally include contributions from the naturally occurring terrestrial radionuclides ({\myKnat}, {\myThnat}, and {\myUnat} with their progeny), which remain present throughout the flight trajectory and constitute a spatially varying background component. Based on the results of the previous hover-flight analysis, no explicit atmospheric radon correction term was included in the present inference model.

The inference approach follows the same MCMC-based posterior sampling applied to the hover flight configurations, but differs in three important respects. First, rather than a single integrated spectrum, inference is performed jointly on the full time series of \qty{1}{\s} pulse-height spectra \mymatht{\mathcal{Y}} with their associated experimental conditions \mymatht{\mathfrak{D}}, which vary continuously as the source--detector separation and relative orientation evolve along the flight trajectory. This requires jointly inferring the naturally occurring terrestrial radionuclide activity concentrations independently for each \qty{1}{\s} spectrum to account for spatially varying background contributions. Second, the position of the anthropogenic sources is included as an additional free parameter and inferred jointly with the source activities and nuisance parameters. Third, because the source mixture is unknown \textit{a priori}, we perform exhaustive Bayesian model comparison over the full power set of {\myCs} and {\myBa} source models, including a background-only model, (see \cref{subsec:fsbi}). For each candidate source model, the inference was restricted to a \qty{6}{\s} temporal window centered on the peak count-rate region.

\begin{figure}[tbh!]
\centering
\includegraphics[width=0.90\textwidth]{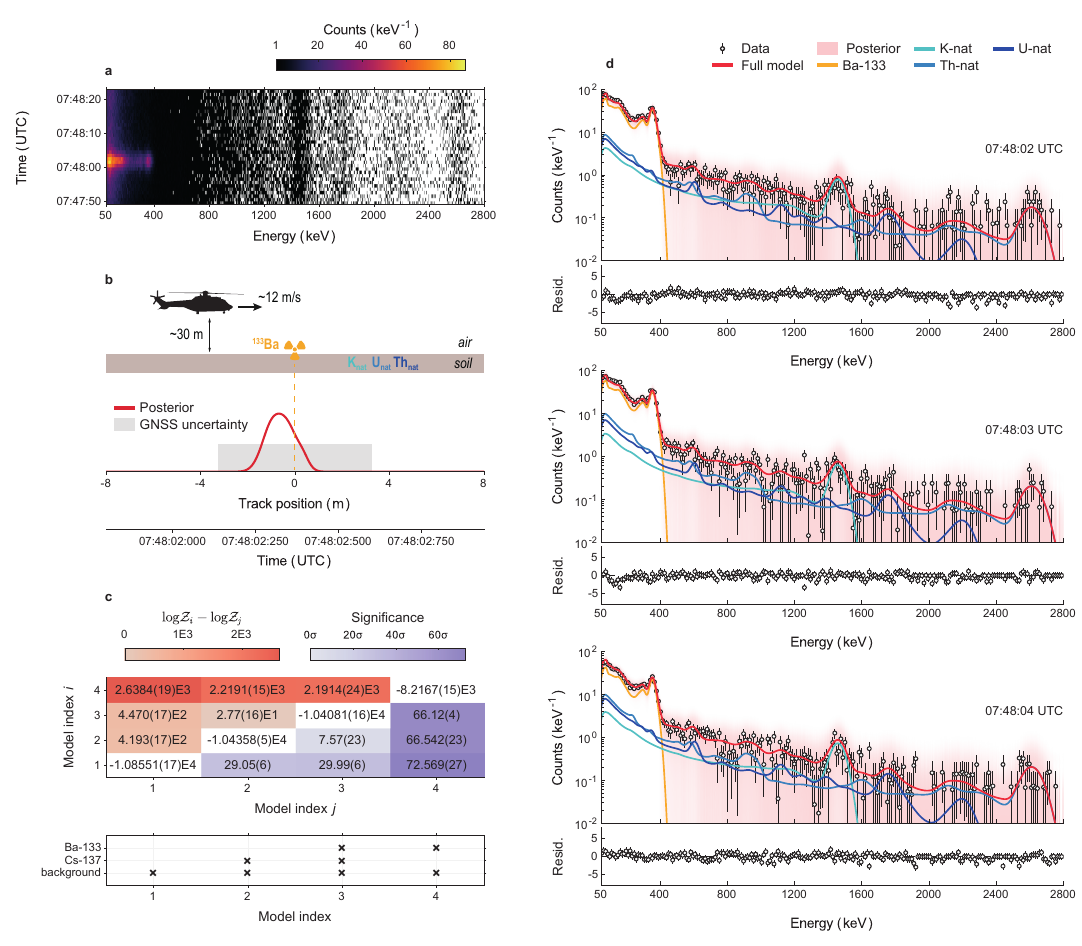}
\caption{\textbf{Bayesian inference on single-pass survey with a single-source {\myBabf} configuration.} \textbf{a}~Spectro-temporal response of the SAGRS system in bin-mode acquisition at \qty{1}{\Hz}. \textbf{b}~Schematic illustrations of the single-pass survey together with the inferred posterior distribution of the anthropogenic source location. \textbf{c}~Bayesian model comparison results. The anti-diagonal entries show the log Bayesian evidence \mymatht{\log\mathcal{Z}_{i=j}} for each source-mixture model (see \cref{subsec:fsbi}). The models are ordered by increasing Bayesian evidence, from the least supported at the bottom left to the most supported at the top right. The entries above the anti-diagonal show the pairwise log Bayes factors \mymathtv{\log\mathcal{B}_{ij}=\log\mathcal{Z}_{i}-\log\mathcal{Z}_{j}}, while those below the anti-diagonal show the corresponding lower bounds on statistical significance, expressed as Gaussian-equivalent significance levels in units of standard deviations \cite{Sellke2001,Trotta2008a}. Related uncertainties are reported as 1~standard errors using least-significant-figure notation. \textbf{d}~Subset of posterior predictive distributions alongside the measured pulse-height spectra for the inferred (true) source mixture with \qty{\sim1}{\s} live time. Statistical uncertainties in the measured pulse-height spectra are shown as 1~standard errors. Residuals are shown as standardized posterior predictive residuals. Posterior predictive distributions are accompanied by point predictions based on the posterior median estimates. Spectral signatures scaled by the posterior median source strengths and measurement live time are also indicated for all sources included in the forward model. For improved interpretability, all spectral quantities were corrected for residual backgrounds, and measured spectra were additionally rebinned from the native \qty{\sim3}{\keV} channel spacing used for the Bayesian inference to an effective bin width of \qty{\sim15}{\keV}. All spectra shown in energy units are derived from calibrated pulse-height channels for visualization only. The line drawing of the AS332M1 helicopter in panel \textbf{b} was adapted from Jetijones, \href{https://creativecommons.org/licenses/by/3.0}{\texttt{CC~BY~3.0}}, via Wikimedia Commons.}
\label{fig:SPBa}
\end{figure}

\begin{figure}[tbh!]
\centering
\includegraphics[width=1\textwidth]{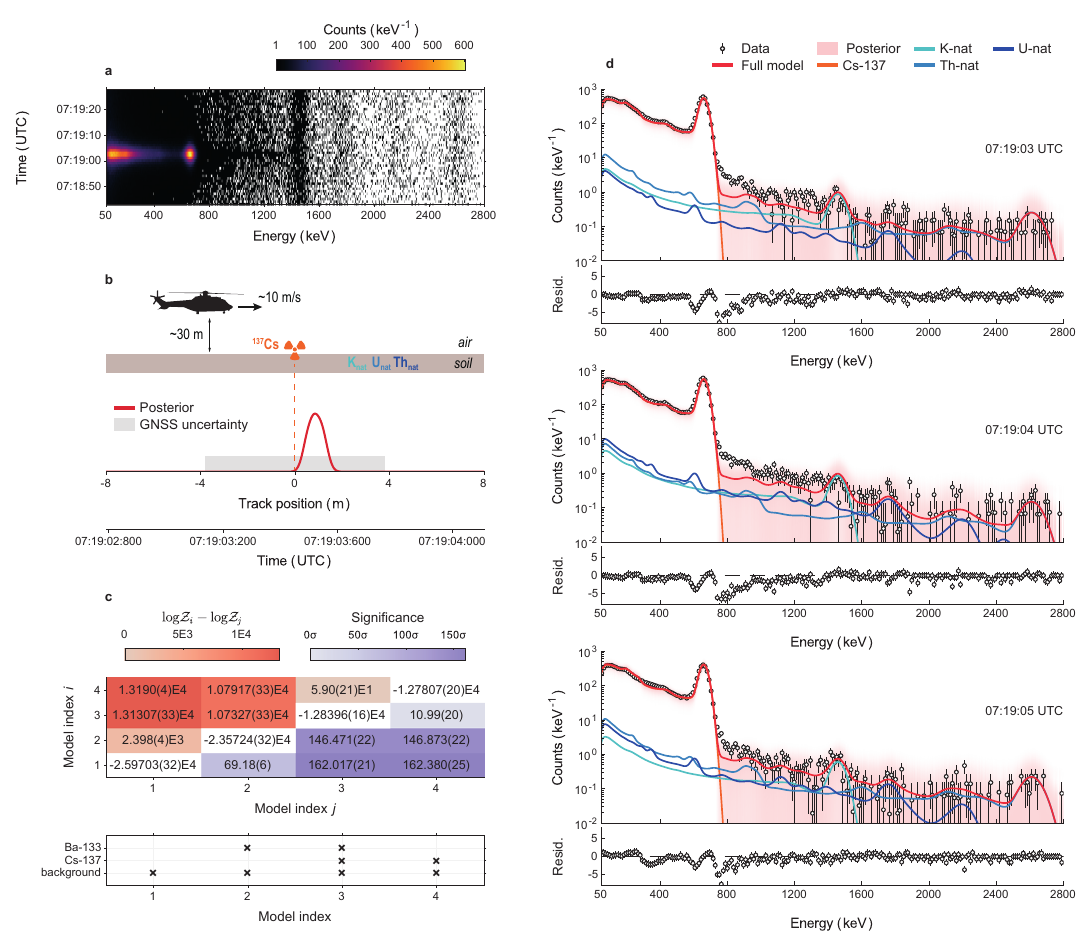}
\caption{\textbf{Bayesian inference on single-pass survey with a single-source {\myCsbf} configuration.} Panels \textbf{a--d} correspond to the same quantities and use the same conventions and processing as in \cref{fig:SPBa}, but for a single-pass survey with an anthropogenic {\myCs} point source deployed instead of {\myBa}. The line drawing of the AS332M1 helicopter in panel \textbf{b} was adapted from Jetijones, \href{https://creativecommons.org/licenses/by/3.0}{\texttt{CC~BY~3.0}}, via Wikimedia Commons.}
\label{fig:SPCs}
\end{figure}

\begin{figure}[tbh!]
\centering
\includegraphics[width=1\textwidth]{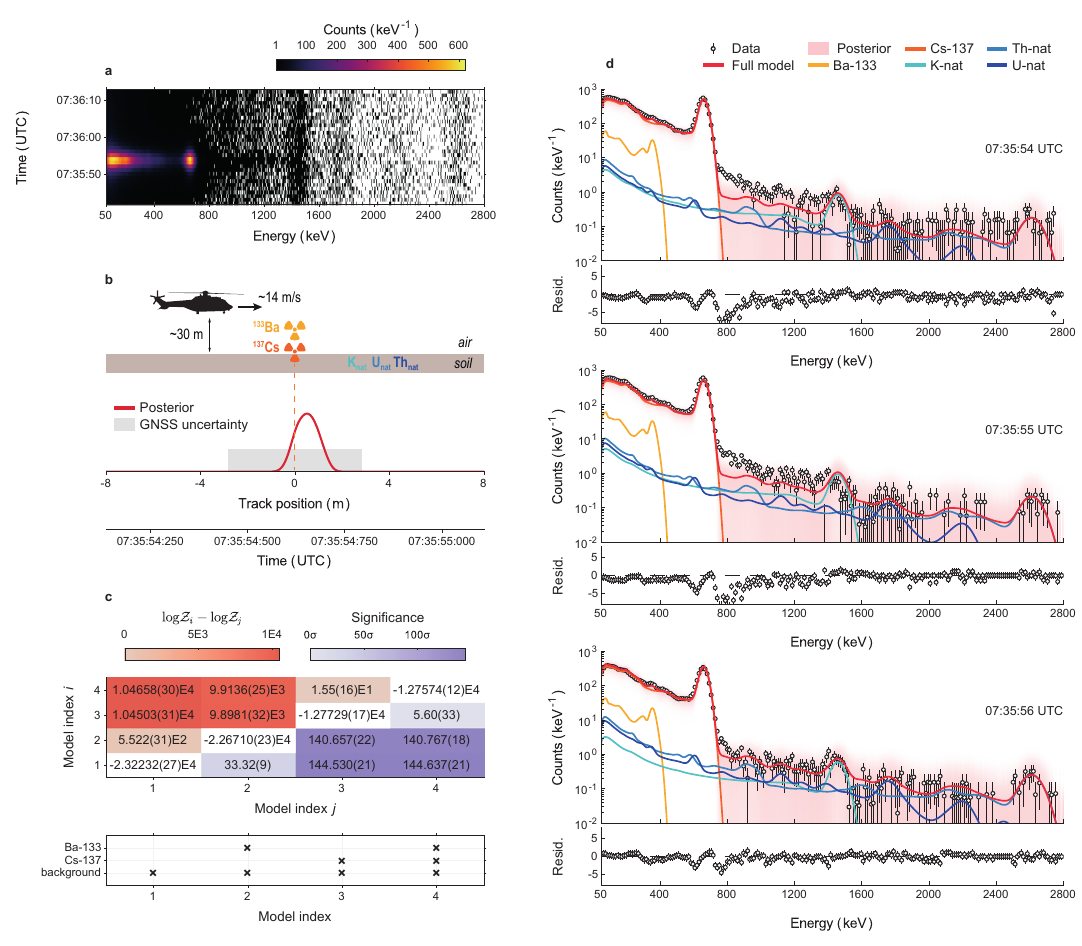}
\caption{\textbf{Bayesian inference on single-pass survey with a dual-source {\myBabf}-{\myCsbf} configuration.} Panels \textbf{a--d} correspond to the same quantities and use the same conventions and processing as in \cref{fig:SPBa}, but for a single-pass survey with combined anthropogenic {\myBa} and {\myCs} point sources deployed instead of a single-source {\myBa} configuration. The line drawing of the AS332M1 helicopter in panel \textbf{b} was adapted from Jetijones, \href{https://creativecommons.org/licenses/by/3.0}{\texttt{CC~BY~3.0}}, via Wikimedia Commons.}
\label{fig:SPBaCs}
\end{figure}

The results for all three survey configurations are presented in \cref{fig:SPBa,fig:SPCs,fig:SPBaCs} and demonstrate three key findings. First, the true source mixture is recovered with decisive statistical evidence in all cases, exceeding the \mymatht{5\sigma} significance threshold relative to the best competing source model (\mymatht{\log\mathcal{B}>15} on Jeffreys' scale \cite{Jeffreys1948}). Second, the inferred source location is recovered with sub-\qty{2}{\m} accuracy, consistent with the onboard global navigation satellite system (GNSS) positioning uncertainty of the aircraft. Third, source activities are retrieved with high accuracy across all configurations. For the single-source deployments, the posterior median activities of \qty{0.473(0.012:0.012)}{\giga\Bq} for {\myBa} and \qty{8.88(0.14:0.12)}{\giga\Bq} for {\myCs} deviate by less than \qty{2}{\percent} from the laboratory reference values. In the dual-source configuration, the inferred activities of \qty{0.49(0.14:0.16)}{\giga\Bq} for {\myBa} and \qty{8.92(0.23:0.21)}{\giga\Bq} for {\myCs} remain within \qty{4}{\percent} of their reference values, with all deviations statistically insignificant within the combined \mymatht{1\sigma} uncertainties. 

Complementing the source inference results, the inferred dispersion parameters \mymatht{\mydispersion} provide a measure of the statistical consistency of the counting process under dynamic survey conditions. For the {\myBa}, {\myCs}, and dual-source {\myBa}-{\myCs} configurations, the posterior median dispersion parameters were inferred as \num{2.8(3.3:1.9)d-3}, \num{2.0(0.4:0.3)d-2}, and \num{2.9(0.6:0.4)d-2}, respectively, all statistically inconsistent with the Poisson limit (\mymatht{\mydispersion=0}). Compared to the corresponding \qty{1}{\s} hover-flight acquisitions, the inferred dispersion parameters are approximately one order of magnitude larger, indicating stronger departures from the idealized Poisson counting model under dynamic flight conditions. However, as discussed in the previous section, elevated values of \mymatht{\mydispersion} cannot be uniquely attributed to intrinsic overdispersion, as they may also reflect residual model bias or other unmodeled effects, such as spatially heterogeneous background fields and detector pileup during high-count-rate source passages. The latter interpretation is supported by the systematic spectral distortions and related elevated residuals observed between \qty{\sim800}{\keV} and \qty{\sim1300}{\keV} in the {\myCs} and dual-source configurations near the peak count-rate instance (\cref{fig:SPCs,fig:SPBaCs}). Importantly, despite the elevated dispersion parameters, the inferred source activities and locations remain statistically consistent with the independent reference measurements within the corresponding credible intervals, indicating robustness of the joint inference under realistic operational conditions.

\section{Discussion}\label{sec:discussion}

A longstanding challenge in mobile gamma-ray spectrometry is the inference of source mixtures, source locations, and source activities from sparse and dynamically acquired survey data. The results presented here show that all three quantities can be recovered jointly within a single full-spectrum Bayesian framework, with percent-level activity accuracy, meter-scale localization accuracy, and decisive source discrimination demonstrated under realistic single-pass survey conditions.

A key computational bottleneck that has so far prevented the practical application of such probabilistic formulations is the need for \myOrdernum{1d4} or more evaluations of the underlying response model for posterior exploration and Bayesian evidence estimation. In realistic survey conditions, the full-spectrum response of an MGRS system depends not only on the candidate source locations, mixtures, and activities, but also on the continuously evolving platform pose as well as environmental conditions. Generating such response estimations directly with high-fidelity Monte Carlo radiation-transport simulations during inference is computationally infeasible for practical survey applications. The framework introduced here overcomes this limitation through a two-stage forward-modeling strategy in which instrument response functions and gamma-ray flux libraries are precomputed on HPC infrastructure, while survey-specific spectral response predictions are generated rapidly during Bayesian inference. This decoupling of radiation transport from statistical inference reduces the computational cost of high-fidelity forward-model evaluations by seven orders of magnitude while preserving the fidelity of the radiation-transport model required for quantitative MGRS over the full spectral range from \qty{\sim50}{\keV} to \qty{\sim3}{\MeV}. As a result, full-spectrum probabilistic inversion and Bayesian model comparison become tractable for realistic mobile gamma-ray survey data acquired under sparse and dynamically evolving measurement conditions.

Beyond computational feasibility, the key practical advance of this framework is the simultaneous inference of source mixtures, source locations, and associated source activities from full-spectrum MGRS data. Unlike conventional FSA pipelines \cite{Minty1992a,Grasty1985a,Minty1998d,Hendriks2001a,Sinclair2011a,Prettyman2006a,Salathe2021a}, the present approach formulates these tasks within a single probabilistic framework. This joint formulation translates into substantial quantitative gains. Across the hover-flight and single-pass experiments, inferred anthropogenic source activities remained statistically consistent with independent laboratory reference values and achieved percent-level agreement even for \qty{1}{\s} acquisition times, representing an order-of-magnitude improvement over previously reported FSA-based approaches under comparable conditions \cite{Ohera2024,Sinclair2016a}. In the single-pass surveys, the correct source mixture was recovered with decisive statistical evidence \mymatht{>\!\!5\sigma} in all tested configurations while source locations and activities were inferred simultaneously. This includes the dual-source {\myBa}-{\myCs} case, where the weaker {\myBa} signal was strongly masked by the dominant {\myCs} contribution, highlighting the ability of the proposed framework to resolve strongly overlapping source signatures that are not separable using traditional peak-based or spectral windowed analyses.

Building on the results presented here, several natural extensions emerge. Our validation focused on terrestrial anthropogenic point sources and natural terrestrial primordial radionuclides distributed in the soil, under both controlled hover-flight conditions and single-pass survey configurations targeting isolated anomalies. Operational radiological surveys, however, often require reconstruction of spatially continuous activity fields over extended areas rather than inference of isolated sources. The present framework can be extended to such problems by embedding the same forward-model and likelihood formulation within a spatially resolved parameterization of the source field, enabling full two- or three-dimensional Bayesian inversion for radiological mapping, contamination assessment, and environmental monitoring. A second important extension concerns the treatment of uncertainty in the known experimental conditions \mymatht{\mathfrak{D}}. In the present work, platform position, orientation, and other measurement-state parameters are treated as deterministic inputs during template generation. In practice, these quantities are themselves subject to measurement uncertainties. Future developments could explicitly propagate these uncertainties through the forward-model generation and Bayesian inference framework, enabling a more complete characterization of model uncertainty and its impact on the inferred source parameters under realistic operational conditions.

More generally, the proposed methodology is both source and platform agnostic. Its scope is not set by the source geometries or helicopter platform used here, but by the ability to construct a physically faithful forward model of the measurement scenario. Because spectral templates are generated numerically from high-fidelity radiation-transport calculations rather than from fixed empirical calibration libraries, the same Bayesian inference protocol can accommodate different source terms and MGRS platform geometries without conceptual modification. This includes, for example, ground-based surveys of distributed soil contamination, maritime surveys involving radionuclides in water or coastal sediments, airborne measurements of gaseous anthropogenic gamma-ray emitters, and spaceborne measurements of secondary cosmic-ray-induced gamma-ray fields on celestial bodies. Importantly, the present study deliberately considers a conservative and computationally demanding scenario involving a crewed airborne platform with characteristic dimensions of \qty{\sim20}{\m} and multiple dynamic platform states that must be accounted for to accurately predict the spectral response \cite{Breitenmoser2025a,Breitenmoser2026}. This represents a near upper bound in forward-model complexity for MGRS applications. Other deployment scenarios are expected to be less demanding, including robotic, UAV-based, and car-borne systems with reduced structural complexity and simpler geometries, as well as spaceborne platforms where atmospheric scattering is absent or strongly reduced. Consequently, we expect the present framework to generalize naturally to a wide range of MGRS platforms and operational environments, provided that sufficiently accurate instrument, platform, and environmental mass models are available.

Taken together, these results demonstrate that full-spectrum Bayesian inference, when coupled with physically faithful forward models, enables accurate and simultaneous estimation of source mixtures, source locations, and source activities under sparse and dynamically evolving measurement conditions, establishing a new level of quantitative capability in mobile gamma-ray spectrometry in terms of both accuracy and joint inferential power. By delivering a complete characterization of the underlying source configuration within a unified inference framework, this approach provides the methodological and computational foundation for a new generation of MGRS analysis pipelines. Building on this foundation, the framework opens the way to a more statistically rigorous and physics-informed era of quantitative mobile gamma-ray spectrometry, enabling reliable decision-making in radiological emergency response, environmental monitoring, nuclear security, and planetary exploration.

\section{Methods}\label{sec:methods}

\subsection{Spectral template computation}\label{subsec:templatemethod}

To overcome the prohibitive computational cost of a single-stage brute-force Monte Carlo approach discussed in \cref{subsec:fsbi}, we adopt a two-stage template generation strategy. As demonstrated in our previous work \cite{Breitenmoser2026}, this method reduces the evaluation cost to \mymatht{\myOrderqty{1d0}{\s}} per template, enabling the rapid generation of high-fidelity spectral template matrices \mymatht{\myvect{M}} required by the probabilistic forward model (see \cref{subsec:bayestheorymethod}). For completeness, we summarize the main elements of this strategy across this and the following two subsections, with additional details available in \myonlinecite{Breitenmoser2026}.

The spectral response of an MGRS system is determined by the convolution of the differential instrument response function \mymatht{\myodvt{R}{E'}(E',E_{\gamma},\myvect{\Upomega}',\myvect{\mathfrak{d}})} with the incident source-strength-normalized double-differential gamma-ray field \mymatht{\mypdvvt{\phi_{\gamma}}{E_{\gamma}}{\varOmega'}(E_{\gamma},\myvect{\Upomega}',\myvect{\mathfrak{d}})}

\begin{align}
     \myodv{\psi}{E'}\left(E',\myvect{\mathfrak{d}}\right) =& %
     \myiintcramp{0}{\infty}{0}{4\pi}\myodv{R}{E'}\left(E',E_{\gamma},\myvect{\Upomega}',\myvect{\mathfrak{d}}\right)\nonumber\\%
     &\times\mypdvv{\phi_{\gamma}}{E_{\gamma}}{\varOmega'}\left(E_{\gamma},\myvect{\Upomega}',\myvect{\mathfrak{d}}\right)%
      \myoodd{\varOmega'}{E_{\gamma}}\label{eq:IRF}%
\end{align}

\noindent with \mymatht{E'}, \mymatht{E_{\gamma}}, \mymatht{\varOmega'}, \mymatht{\myvect{\Upomega}'}, and \mymatht{\myvect{\mathfrak{d}}} denoting the spectral energy related to the pulse-height in the gamma-ray spectrometer, the energy of the incident gamma ray, the solid angle and direction unit vector in the local non-inertial platform-fixed coordinate system, as well as the known experimental condition vector, respectively. In the present work, \mymatht{\myvect{\mathfrak{d}}} comprises the dynamic helicopter pose relative to the ground, fuel state, and ambient environmental conditions (air temperature, pressure, and humidity), as introduced in \cref{subsec:fsbi}. For computational evaluation, the integral in \cref{eq:IRF} is discretized over spectral energy and direction, allowing us to recast \cref{eq:IRF} in matrix notation as

\begin{equation}
    \myvect{\uppsi}\left(\myvect{\mathfrak{d}}\right) \approx  
    \sum_{k=1}^{N_{\myvect{\Upomega}'}} \myvect{R}\left(\myvect{\Upomega}'_k, \myvect{\mathfrak{d}}\right) \myvect{\upphi}_{\gamma}\left(\myvect{\Upomega}'_k, \myvect{\mathfrak{d}}\right) \label{eq:IRFm}
\end{equation}

\noindent where \mymatht{\myvect{\uppsi} \in \myRealul{N_y}{+}} denotes the spectral template vector, \mymathtv{\myvect{R} \in \myRealul{N_y \times N_{E_{\gamma}}}{+}} the instrument response matrix, and \mymathtv{\myvect{\upphi}_{\gamma} \in \myRealul{N_{E_{\gamma}}}{+}} the gamma-ray flux vector. The individual elements of the instrument response matrix are estimated by scaling the differential instrument response components with the corresponding spectral energy bin width, i.e., \mymathtv{R(E',E_{\gamma},\myvect{\Upomega}'_k, \myvect{\mathfrak{d}}) \approx \myodvt{R}{E'}(E',E_{\gamma}, \myvect{\Upomega}'_k, \myvect{\mathfrak{d}})\myincrement{E'}}. Similarly, we compute the individual elements of the gamma-ray flux vector by scaling the double-differential gamma-ray flux components with the corresponding energy and solid angle bin widths, i.e., \mymathtv{\phi_{\gamma}(E_{\gamma}, \myvect{\Upomega}'_k, \myvect{\mathfrak{d}}) \approx \mypdvvt{\phi_{\gamma}}{E_{\gamma}}{ \varOmega'}(E_{\gamma}, \myvect{\Upomega}'_k, \myvect{\mathfrak{d}}) \myincrement{E_{\gamma}}\myincrement{\varOmega'_k}}.

To compute the spectral templates \mymatht{\myvect{\uppsi}} as defined in \cref{eq:IRFm}, we implemented the convolution using the \mycode{pagemtimes} routine in \mycode{MATLAB\textsuperscript{\textregistered}} (Version \myversion{R2024a}), which enables efficient multithreaded evaluation of the directional matrix-vector products. Prior to convolution, the IRF and flux libraries must be interpolated onto a common discretization scheme. Spectral alignment is achieved by interpolating the IRF onto the regular energy grid of the gamma-ray flux simulation \citep{Jandel2004a}, while angular alignment is handled by azimuthal expansion in \qty{30}{\degree} steps, followed by mapping to the local platform-fixed frame using Tait-Bryan rotations that define the platform's instantaneous attitude relative to the global frame. The aligned quantities are then convolved to produce the spectral templates, with statistical and systematic uncertainties propagated through the pipeline. On a standard workstation (8 cores, \qty{2.1}{\GHz}), the total runtime for template generation including uncertainty estimation is \myOrderqty{1}{\s} per template. In the next subsections, we detail the generation of both the instrument response matrix and the gamma-ray flux vector libraries for the benchmark studies discussed in \cref{subsec:hfresults,subsec:spresults}.

\subsection{Instrument response function}\label{subsec:irfmethod}

The IRFs introduced in the previous section were estimated using high-fidelity Monte Carlo radiation transport simulations of the validated SAGRS mass model \cite{Breitenmoser2025a}. Following established practices for spaceborne gamma-ray spectrometers \cite{Prettyman2006a,Kobayashi2010a,Peplowski2011b,Prettyman2011a}, we illuminated the system with circular monoenergetic plane waves over a grid of gamma-ray energies and incident directions in the platform-fixed coordinate system. The IRF elements were derived as the normalized spectral counts in each pulse-height channel, with source radii chosen to ensure invariance of the response (see \myonlinecite{Breitenmoser2026}). The simulation grid spanned \num{30} energies between \qty{50}{\keV} and \qty{3}{\MeV} combined with \num{134} incident directions, yielding \num{4020} runs. To account for temporal variability of the platform, we generated IRFs for eight distinct mass-model states defined by fuel load, crew occupancy, and landing gear position, including those states encountered during the survey described in \cref{subsec:dataresult}. Each simulation tracked \num{1d7} primary histories using \mycode{FLUKA} (v4-2.2) \cite{Ahdida2022a} with the \mycode{precisio} physics settings, providing coupled photon-electron-positron transport, secondary particle production, and X-ray fluorescence. To reproduce the nonproportional scintillation response of the SAGRS system, a dedicated scintillation model calibrated in prior work \cite{Breitenmoser2023c,Breitenmoser2025a} was implemented through \mycode{comscw} and \mycode{usrglo}. Motivated by the range of the transported particles, transport thresholds were set to \qty{1}{\keV} in the detector volume and \qty{10}{\keV} elsewhere. Energy deposition events were recorded per crystal using custom \mycode{usreou} routines and converted into expected pulse-height spectra using the \mycode{NPScinMC} pipeline, leveraging energy and resolution calibration models as detailed in \myonlinecite{Breitenmoser2025a} (see also \cref{subsec:dataresult}). Due to the substantial computational cost of these Monte Carlo simulations, we employed HPC infrastructure available at the Paul Scherrer Institute (\qty{2.6}{\giga\Hz} nominal clock speed). Generating the complete set of IRFs required \qty{\sim5.5d4}{\corehour}. These precomputed IRFs represent a one-time computational investment for a given detector and platform configuration and can subsequently be reused across multiple template-generation and Bayesian inference tasks, thereby avoiding repeated high-fidelity radiation-transport simulations while preserving the accuracy of the underlying forward model.

\subsection{Gamma-ray flux library}\label{subsec:fluxmethod}

We generated double-differential gamma-ray flux libraries for all sources included in the forward models using the \mycode{FLUKA} code (v4-2.2) \cite{Ahdida2022a} on HPC infrastructure. Specifically, we adopted the built-in \mycode{USRYIELD} method to score the double-differential flux with a bin width of \qty{0.5}{\keV} for the gamma-ray energy \mymatht{E_{\gamma}} and \qty{5}{\degree} for the polar angle \mymatht{\theta} in a global inertial coordinate system (exploiting azimuthal invariance). The double-differential flux was evaluated on a grid of ground clearances from \qty{10}{\m} to \qty{250}{\m} in \qty{10}{\m} increments using the reciprocal transform method \cite{Baldoncini2018b,Kastlander2005a,Namito2012a}. The resulting flux libraries were subsequently interpolated to the recorded platform ground clearance for each spectrum instance during spectral template generation (see also \cref{subsec:bayestheorymethod}). Similar to the IRF generation, we adopted the \mycode{precisio} physics settings with a \qty{10}{\keV} transport threshold. The accumulated computation time was \qty{\sim1d2}{\corehour} per source, with \num{2d9} primary particle histories, where each history corresponds to a full Monte Carlo transport cascade initiated by a single radionuclide decay according to the predefined source emission model.

For all flux simulations, we adopted a dynamic environmental mass model with three main mass elements, i.e., the atmosphere, the terrestrial soil, as well as paved surfaces present at the training ground. These components were modeled as homogeneous media with compositional data sourced from \myonlinecite{McConn2011a}. The atmosphere was modeled for each measurement separately as homogeneous humid air as a function of the recorded air temperature, pressure, and humidity (see Supplementary Table~1). Motivated by the mean free path of the primary photons in air, atmospheric boundaries were confined to a \qty{3}{\km} radius sphere, centered at the SAGRS location. The ground was limited to a vertical extension of \qty{10}{\m}. In addition to the environmental media, we included also detailed mass models of the deployed radionuclide sources and custom source holders (see \myonlinecite{Breitenmoser2025a}). 

Radionuclide emissions were simulated with the \mycode{FLUKA} \mycode{raddecay} module in semi-analogue mode, according to related decay schemes. Anthropogenic source volumes were explicitly modeled, with their source strengths parameterized by the absolute source activity in \unit{\Bq}. For the natural terrestrial radionuclides ({\myKnat}, {\myThnat}, {\myUnat}), we assumed secular equilibrium with progeny and a spatially uniform distribution within the soil. Corresponding source strengths were parameterized by the activity mass concentration in \unit{\Bq\per\kg}. Similarly, the atmospheric radon source term discussed in \cref{subsec:hfresults} was implemented as a homogeneous atmospheric {\myRnIsotope} volume source in secular equilibrium with its progeny up to but excluding \mynuclt{210}{82}{Pb}, and with the source strength  expressed as activity volume concentration in \unit{\Bq\per\cubic\m}.

\subsection{Swiss Airborne Gamma-Ray Spectrometry System}\label{subsec:sagrs}

The Swiss Airborne Gamma-Ray Spectrometry (SAGRS) system is a helicopter-borne MGRS platform jointly operated by the National Emergency Operations Centre (NEOC), the Nuclear Biological Chemical defence and Explosive Ordnance Disposal Centre of Excellence (NBC-EOD), and the Swiss Air Force for environmental surveys and emergency response. The system is fully deployable and airborne within four hours, enabling rapid and flexible operations under a wide range of survey and emergency scenarios. A brief technical summary is provided below. Comprehensive specifications and validation studies are reported in \myonlinecitepl{Breitenmoser2022,Breitenmoser2025a,Breitenmoser2026}.

The detector assembly consists of four prismatic NaI(Tl) scintillation crystals (Saint-Gobain 4×4H16/3.5-X), each measuring \mymathtv{\qty{10.2}{\cm}\times\qty{10.2}{\cm}\times\qty{40.6}{\cm}}. The crystals are individually enclosed in aluminum housings, optically coupled to Hamamatsu R10755 photomultiplier tubes, and equipped with dedicated readout electronics. They are embedded in thermally insulating, vibration-damping polyethylene foam and mounted in a rugged aluminum box (outer dimensions \mymathtv{\qty{90}{\cm}\times\qty{64}{\cm}\times\qty{35}{\cm}}). The unit is installed in the cargo bay of an A\'{e}rospatiale AS332M1 Super Puma helicopter, aligned with the fuselage underside to maximize sensitivity to terrestrial radiation. System integration includes avionics via an ARINC 429 bus. Recorded parameters comprise GNSS position, radar altitude, helicopter orientation, and fuel state. These auxiliary measurements are synchronized with the spectrometer data and serve as input to the forward model. Survey operations are controlled via two onboard operator stations interfaced with a central data server, with the complete system, including support hardware, having a total mass of \qty{\sim290}{\kg}.

Spectral acquisition is performed in bin mode at \qty{1}{\Hz}, with individual spectra recorded for each of the four scintillator assemblies, alongside a combined sum spectrum representing the aggregated detector response. Each spectrum comprises \num{1008} channels spanning \qty{\sim50}{\keV} to \qty{\sim3}{\MeV}. The readout electronics provide automatic gain stabilization by leveraging spectral information from natural potassium and thorium gamma-ray emission lines together with crystal temperature measurements from auxiliary thermocouples. In addition, the system performs spectral linearization with offset correction and live-time tracking. These procedures ensure a stable and approximately linear mapping between pulse-height channels and energy over the duration of a measurement campaign. Consequently, all Bayesian inference is performed in pulse-height channel space, which provides a stable and detector-native representation of the measured spectra.

\subsection{Probabilistic forward modeling}\label{subsec:bayestheorymethod}

Following the Bayesian framework outlined in \cref{subsec:fsbi}, we adopt a negative binomial probabilistic forward model to describe the observed counts \mymatht{C\in\myNat} in the set of recorded pulse-height spectra \mymatht{\mathcal{Y}}, accounting for both measurement uncertainty and overdispersion relative to the ideal Poisson model (see \cref{sec:Introduction}). The associated likelihood in \cref{eq:BayesTheorem} is formulated using a parameterization established in prior studies \cite{Hunnefeld2022a,Salinas2020a,Lloyd-Smith2007a}, where the dispersion parameter \mymatht{\mydispersion \in \myReall{+}} relaxes the Poisson variance constraint to
\mymatht{\operatorname{Var}(\myrandomvec{Y})=\langle\myrandomvec{Y}\rangle+\mydispersion\langle\myrandomvec{Y}\rangle^2}, with \mymatht{\myrandomvec{Y}} denoting the random vector associated with the pulse-height spectrum {\myvect{y}} (see \cref{subsec:fsbi}):

\begin{equation}
    \mathcal{L}\left(\myvect{\uptheta};\mathcal{Y},\mathfrak{D},\mathcal{S}\right) = \myprod{k=1}{N_{\mathcal{Y}}}\myprod{j=1}{N_{y}} \frac{\mygammaf{C_{k,j}+\frac{1}{{\mydispersion}}}}{\mygammaf{\frac{1}{{\mydispersion}}}\mygammaf{C_{k,j}+1}}\left(\frac{1}{1+{\mydispersion}\mathcal{M}(\Xi,\mathcal{X},\myvect{\mathfrak{d}}_{k})}\right)^{\frac{1}{{\mydispersion}}}
    \left(\frac{1}{1+[{\mydispersion}\mathcal{M}(\Xi,\mathcal{X},\myvect{\mathfrak{d}}_{k})]^{-1}}\right)^{C_{k,j}}.\label{eq:likeli}
\end{equation}

\noindent Here, \mymatht{\mygammaf{\cdot}} denotes the gamma function and \mymatht{\mathcal{M}{(\cdot})} the deterministic forward operator, which maps source strengths \mymatht{\Xi=\{\xi_s\in\myReall{+}\}^{N_s}_{s=1}}, source positions \mymatht{\mathcal{X}=\{\myvect{x}_s\in\myRealu{3}\}^{N_s}_{s=1}}, and (known) experimental conditions \mymatht{\mathfrak{D}=\{\myvect{\mathfrak{d}}_t \in \myRealu{N_\mathfrak{d}}\}^{N_\mathfrak{D}}_{t=1}} for a given source mixture \mymatht{\mathcal{S}} containing \mymatht{N_s\in\myNat} sources to expected counts, with \mymatht{C_{k,j}} denoting the observed counts in pulse-height channel \mymatht{j} of spectrum \mymatht{k} (see \cref{subsec:fsbi}). In the present negative binomial formulation, the joint parameter vector \mymatht{\boldsymbol{\uptheta}} is formed by stacking the source strengths \mymatht{\Xi}, source locations \mymatht{\mathcal{X}}, and the dispersion parameter \mymatht{\mydispersion}. The experimental conditions \mymatht{\mathfrak{D}} are evaluated (or interpolated where required) at the corresponding spectrum instances \mymatht{k = 1,\ldots,N_{\mathcal{Y}}} via the mapping \mymatht{\myvect{\mathfrak{d}}_k \leftrightarrow t_k}, thereby linking each measured spectrum to its associated acquisition conditions. Following the FSA approach \cite{Paradis2020a,Andre2021a}, we define \mymatht{\mathcal{M}} as the linear superposition of spectral templates \mymatht{\myvect{\uppsi}\in\myRealul{N_y}{+}}  (see \cref{subsec:templatemethod}) during the survey with measurement live time  \mymatht{\myincrement{t}_{\mathrm{gr}}}:

\begin{equation}
    \mathcal{M}\left(\Xi,\mathcal{X},\myvect{\mathfrak{d}}\right)
    = \myintcramp{0}{\myincrement{t}_{\mathrm{gr}}}\myvect{M}\left(\mathcal{X},\myvect{\mathfrak{d}}\right)\myvect{\xi}+\myvect{c}_\text{b}\left(\myvect{\mathfrak{d}}\right)\myodd{t}\label{eq:ForwardModelTheory}
\end{equation}

\noindent where \mymathtv{\myvect{M}\in\myRealul{N_y \mytimes N_s}{+}} denotes the spectral template matrix \mymathtv{\myvect{M}(\mathcal{X},\myvect{\mathfrak{d}}) \coloneqq[ \myvect{\uppsi}^{(1)}(\myvect{x}_1,\myvect{\mathfrak{d}}),\allowbreak \myvect{\uppsi}^{(2)}(\myvect{x}_2,\myvect{\mathfrak{d}}),\cdots,\allowbreak \myvect{\uppsi}^{(N_s)}(\myvect{x}_{N_s},\myvect{\mathfrak{d}})]}, together with \mymatht{\myvect{\xi}\in\myRealul{N_s}{+}} specifying the joint source-strength vector formed by stacking the source strengths contained in \mymatht{\Xi}, and \mymatht{\myvect{c}_\text{b}\in\myRealul{N_y}{+}} encoding the spectral response to known gamma-ray backgrounds. 

It is worth adding that, because the gamma function terms in \cref{eq:likeli} grow rapidly for moderately large arguments, special care is required when numerically evaluating the likelihood. To avoid under- or overflow, it is advisable to operate with the logarithm of the likelihood function, known as the log-likelihood, instead of directly evaluating \cref{eq:likeli}:

\begingroup
\allowdisplaybreaks
\begin{multline}
\log{\mathcal{L}\left(\myvect{\uptheta};\mathcal{Y},\mathfrak{D},\mathcal{S}\right)} = \mysum{k=1}{N_{\mathcal{Y}}}\mysum{j=1}{N_{y}}\log{\mygammaf{C_{k,j}+\frac{1}{{\mydispersion}}}}-\log{\mygammaf{\frac{1}{{\mydispersion}}}}\\%
    -\log{\mygammaf{C_{k,j}+1}}-\frac{1}{{\mydispersion}}\log{\left(1+{\mydispersion}\mathcal{M}(\Xi,\mathcal{X},\myvect{\mathfrak{d}}_{k})\right)}\\%
    -C_{k,j}\log{\left(1+\frac{1}{{\mydispersion}\mathcal{M}(\Xi,\mathcal{X},\myvect{\mathfrak{d}}_{k})}\right)}\label{eq:loglikeli}
\end{multline}
\endgroup

\noindent with numerically stable implementations of the log-gamma function \mymatht{\log\mygammaf{\cdot}} being readily available in all major scientific computing environments, including \mycode{Python}, \mycode{MATLAB\textsuperscript{\textregistered}}, and \mycode{C++}.

The likelihood specification in \cref{eq:likeli} forms the first element of the Bayesian framework. Completing the posterior model in \cref{eq:BayesTheorem} requires assigning suitable priors to the model parameters \mymatht{\myvect{\uptheta}}. To avoid overly restrictive assumptions on these prior distributions, we adopted weakly informative, statistically independent marginal priors \mymathtv{\pi\left(\myvect{\uptheta}\mid\mathcal{S}\right) \coloneqq \prod_{m=1}^{M}\pi\left( \theta_{m}\mid\mathcal{S}\right)} for all $M$ model parameters \mymatht{\theta} (see also \cref{subsec:fsbi}). For parameters that are constrained to half-spaces, in particular the source strengths and the dispersion parameter, we adopt truncated univariate normal distributions with support restricted to the physically admissible domain. For parameters defined on finite bounded domains, such as the source positions in single-pass survey configurations, we employ uniform priors, reflecting the absence of prior spatial preference within the measurement area. A full list of all marginal prior distributions is provided in Supplementary Table~2.

\subsection{Bayesian computation}\label{subsec:bayescompmethod}

All Bayesian computations were performed with the \mycode{UQLab} code \cite{Marelli2014a} using an affine invariant ensemble MCMC algorithm \cite{Goodman2010} to sample the posterior distribution \mymatht{\pi\left(\myvect{\uptheta}\mid \mathcal{Y},\mathfrak{D},\mathcal{S}\right)} in \cref{eq:BayesTheorem} via Bayes' theorem combined with an MCMC-based marginal likelihood estimator \cite{Perrakis2014} to evaluate the Bayesian evidence \mymatht{\mathcal{Z}}. Lower bounds on the statistical significance reported in \cref{subsec:spresults} were derived from Bayes factors by conversion to equivalent Gaussian $\sigma$-levels using conservative calibrations between p-values and Bayes factors \cite{Sellke2001,Trotta2008a}. The computational cost of the Bayesian inversion remained moderate, with a complete MCMC run requiring \myOrderqty{1d2}{\s} wall-clock time on a single Intel Xeon Gold 6130 processor core (\qty{2.1}{\giga\Hz}). The convergence and precision of the MCMC simulations were carefully assessed using standard diagnostic tools \cite{Brooks1998a,Gelman2013}, showing a potential scale reduction factor \mymatht{\hat{R}<\num{1.02}} and effective sample size \mymatht{\text{ESS}>\num{600}} across all MCMC runs.  Additional trace and convergence plots for the individual parameters \mymatht{\theta} and point estimators (Supplementary Figs.~2--6) alongside detailed Bayesian inversion summaries (Supplementary Tables~3 and 4) can be found in the Supplementary Information File for this study.

\backmatter

\bmhead{Data availability}
All experimental and template data presented herein have been deposited in the Zenodo repository under accession code \url{https://doi.org/10.5281/zenodo.18004440} \citep{Breitenmoser2025k}.

\bmhead{Code availability}
The \mycode{FLUKA} code \citep{Ahdida2022a} used for Monte Carlo radiation transport and detector response simulations is available at \url{https://fluka.cern/}. We adopted the graphical user interface \mycode{FLAIR} \citep{Vlachoudis2009a}, available at \url{https://flair.web.cern.ch/flair/}, to set up the \mycode{FLUKA} input files and create the mass model figures. The custom \mycode{FLUKA} user routines adopted in the Monte Carlo simulations have been deposited in the ETH Research Collection repository under accession code \url{https://doi.org/10.3929/ethz-b-000595727} \citep{Breitenmoser2023}. Data processing and Bayesian inference computation were performed using the $\mycode{MATLAB}\textsuperscript{\textregistered}$ code in combination with the open-source toolbox \mycode{UQLab} \citep{Marelli2014a}, available at \url{https://www.uqlab.com/}. Figures were created using $\mycode{MATLAB}\textsuperscript{\textregistered}$ and the vector graphics editor \mycode{Adobe Illustrator}.

\bmhead{Acknowledgements}
\noindent We gratefully acknowledge the support by the members of the National Emergency Operations Centre (NEOC), the Swiss Armed Forces, specifically the Swiss Air Force and the Nuclear, Biological, Chemical, Explosive Ordnance Disposal and Mine Action Centre of Excellence (NBC-EOD), as well as the Expert Group Airborne Gamma Spectrometry for their support in conducting the radiation measurements described in this study. Our sincere thanks go to Gernot Butterweck for his invaluable scientific expertise and support during the measurements, as well as for his role in supervision. Finally, we extend our gratitude to Dominik Werthmüller for his technical support in running the Monte Carlo simulations on the computer cluster at the Paul Scherrer Institute.

\bmhead{Funding}
\noindent This research was partially supported by the Swiss Federal Nuclear Safety Inspectorate (grant no. CTR00836 \& CTR00491).

\bmhead{Author contributions}
\noindent D.B. conceptualized the work, designed the methodology, performed the investigations, curated and analyzed the data, developed the software, validated and visualized the results, and supervised the project. D.B. wrote the original draft of the manuscript, and together with A.S., M.M.K. and S.M. reviewed and edited it. A.S. contributed to the investigations. S.M. supervised the project and acquired funding.

\bmhead{Competing interests}
The authors declare no competing interests.

\bmhead{Additional information}
The online version contains supplementary material.


\begin{thebibliography}{100}
\expandafter\ifx\csname url\endcsname\relax
  \def\url#1{\burl{#1}}\fi
\expandafter\ifx\csname urlprefix\endcsname\relax\def\urlprefix{URL }\fi
\providecommand{\bibinfo}[2]{#2}
\providecommand{\eprint}[2][]{\url{#2}}
\providecommand{\doi}[1]{\url{https://doi.org/#1}}
\bibcommenthead

\bibitem{Jones2001a}
\bibinfo{author}{Jones, D.~G.}
\newblock \bibinfo{title}{Development and application of marine gamma-ray measurements: {{A}} review}.
\newblock \emph{\bibinfo{journal}{J. Environ. Radioact.}} \textbf{\bibinfo{volume}{53}}, \bibinfo{pages}{313--333} (\bibinfo{year}{2001}).

\bibitem{Lee2023c}
\bibinfo{author}{Lee, M.~S.} \emph{et~al.}
\newblock \bibinfo{title}{Real-time wireless marine radioactivity monitoring system using a {{SiPM-based}} mobile gamma spectroscopy mounted on an unmanned marine vehicle}.
\newblock \emph{\bibinfo{journal}{Nucl. Eng. Technol.}} \textbf{\bibinfo{volume}{55}}, \bibinfo{pages}{2158--2165} (\bibinfo{year}{2023}).

\bibitem{Rosenthal1991a}
\bibinfo{author}{Rosenthal, J.~J.}, \bibinfo{author}{{de Almeidat}, C.~E.} \& \bibinfo{author}{Mendonca, A.~H.}
\newblock \bibinfo{title}{The {{Radiological Accident}} in {{Goiania}}}.
\newblock \emph{\bibinfo{journal}{Health Phys.}} \textbf{\bibinfo{volume}{60}}, \bibinfo{pages}{7--15} (\bibinfo{year}{1991}).

\bibitem{Drovnikov1997a}
\bibinfo{author}{Drovnikov, V.~V.}, \bibinfo{author}{Egorov, N.~Y.}, \bibinfo{author}{Kovalenko, V.~V.}, \bibinfo{author}{Serboulov, Y.~A.} \& \bibinfo{author}{Zadorozhny, Y.~A.}
\newblock \bibinfo{title}{Some results of the airborne high energy resolution gamma-spectrometry application for the research of the {{USSR European}} territory radioactive contamination in 1986 caused by the {{Chernobyl}} accident}.
\newblock \emph{\bibinfo{journal}{J. Environ. Radioact.}} \textbf{\bibinfo{volume}{37}}, \bibinfo{pages}{223--234} (\bibinfo{year}{1997}).

\bibitem{Sanada2014a}
\bibinfo{author}{Sanada, Y.}, \bibinfo{author}{Sugita, T.}, \bibinfo{author}{Nishizawa, Y.}, \bibinfo{author}{Kondo, A.} \& \bibinfo{author}{Torii, T.}
\newblock \bibinfo{title}{The aerial radiation monitoring in {{Japan}} after the {{Fukushima Daiichi}} nuclear power plant accident}.
\newblock \emph{\bibinfo{journal}{Nucl. Sci. Technol.}} \textbf{\bibinfo{volume}{4}}, \bibinfo{pages}{76--80} (\bibinfo{year}{2014}).

\bibitem{Deal1972a}
\bibinfo{author}{Deal, L.~J.}, \bibinfo{author}{Doyle, J.~F.}, \bibinfo{author}{Burson, Z.~G.} \& \bibinfo{author}{Boyns, P.~K.}
\newblock \bibinfo{title}{Locating the lost athena missile in mexico by the aerial radiological measuring system ({{ARMS}})}.
\newblock \emph{\bibinfo{journal}{Health Phys.}} \textbf{\bibinfo{volume}{23}}, \bibinfo{pages}{95--98} (\bibinfo{year}{1972}).

\bibitem{Hellfeld2021a}
\bibinfo{author}{Hellfeld, D.} \emph{et~al.}
\newblock \bibinfo{title}{Free-moving {{Quantitative Gamma-ray Imaging}}}.
\newblock \emph{\bibinfo{journal}{Sci. Rep.}} \textbf{\bibinfo{volume}{11}}, \bibinfo{pages}{20515} (\bibinfo{year}{2021}).

\bibitem{Curtis2020a}
\bibinfo{author}{Curtis, J.~C.} \emph{et~al.}
\newblock \bibinfo{title}{Simulation and validation of the {{Mobile Urban Radiation Search}} ({{MURS}}) gamma-ray detector response}.
\newblock \emph{\bibinfo{journal}{Nucl. Instrum. Methods Phys. Res. Sect. Accel. Spectrometers Detect. Assoc. Equip.}} \textbf{\bibinfo{volume}{954}}, \bibinfo{pages}{161128} (\bibinfo{year}{2020}).

\bibitem{Salathe2021a}
\bibinfo{author}{Salathe, M.} \emph{et~al.}
\newblock \bibinfo{title}{Determining urban material activities with a vehicle-based multi-sensor system}.
\newblock \emph{\bibinfo{journal}{Phys. Rev. Res.}} \textbf{\bibinfo{volume}{3}}, \bibinfo{pages}{023070} (\bibinfo{year}{2021}).

\bibitem{Fishman1994}
\bibinfo{author}{Fishman, G.~J.} \emph{et~al.}
\newblock \bibinfo{title}{Discovery of {{Intense Gamma-Ray Flashes}} of {{Atmospheric Origin}}}.
\newblock \emph{\bibinfo{journal}{Science}} \textbf{\bibinfo{volume}{264}}, \bibinfo{pages}{1313--1316} (\bibinfo{year}{1994}).

\bibitem{Tavani2011}
\bibinfo{author}{Tavani, M.} \emph{et~al.}
\newblock \bibinfo{title}{Terrestrial {{Gamma-Ray Flashes}} as {{Powerful Particle Accelerators}}}.
\newblock \emph{\bibinfo{journal}{Phys. Rev. Lett.}} \textbf{\bibinfo{volume}{106}}, \bibinfo{pages}{018501} (\bibinfo{year}{2011}).

\bibitem{Smith2011a}
\bibinfo{author}{Smith, D.~M.} \emph{et~al.}
\newblock \bibinfo{title}{A terrestrial gamma ray flash observed from an aircraft}.
\newblock \emph{\bibinfo{journal}{J. Geophys. Res. Atmospheres}} \textbf{\bibinfo{volume}{116}}, \bibinfo{pages}{20124} (\bibinfo{year}{2011}).

\bibitem{Neubert2020a}
\bibinfo{author}{Neubert, T.} \emph{et~al.}
\newblock \bibinfo{title}{A terrestrial gamma-ray flash and ionospheric ultraviolet emissions powered by lightning}.
\newblock \emph{\bibinfo{journal}{Science}} \textbf{\bibinfo{volume}{367}}, \bibinfo{pages}{183--186} (\bibinfo{year}{2020}).

\bibitem{Sinclair2011a}
\bibinfo{author}{Sinclair, L.~E.} \emph{et~al.}
\newblock \bibinfo{title}{Aerial measurement of radioxenon concentration off the west coast of vancouver island following the fukushima reactor accident}.
\newblock \emph{\bibinfo{journal}{J. Environ. Radioact.}} \textbf{\bibinfo{volume}{102}}, \bibinfo{pages}{1018--1023} (\bibinfo{year}{2011}).

\bibitem{Baldoncini2017a}
\bibinfo{author}{Baldoncini, M.} \emph{et~al.}
\newblock \bibinfo{title}{Exploring atmospheric radon with airborne gamma-ray spectroscopy}.
\newblock \emph{\bibinfo{journal}{Atmos. Environ.}} \textbf{\bibinfo{volume}{170}}, \bibinfo{pages}{259--268} (\bibinfo{year}{2017}).

\bibitem{Prettyman2006a}
\bibinfo{author}{Prettyman, T.~H.} \emph{et~al.}
\newblock \bibinfo{title}{Elemental composition of the lunar surface: {{Analysis}} of gamma ray spectroscopy data from {{Lunar Prospector}}}.
\newblock \emph{\bibinfo{journal}{J. Geophys. Res. Planets}} \textbf{\bibinfo{volume}{111}} (\bibinfo{year}{2006}).

\bibitem{Hahn2007a}
\bibinfo{author}{Hahn, B.~C.} \emph{et~al.}
\newblock \bibinfo{title}{Mars {{Odyssey Gamma Ray Spectrometer}} elemental abundances and apparent relative surface age: {{Implications}} for {{Martian}} crustal evolution}.
\newblock \emph{\bibinfo{journal}{J. Geophys. Res. Planets}} \textbf{\bibinfo{volume}{112}} (\bibinfo{year}{2007}).

\bibitem{Kobayashi2010a}
\bibinfo{author}{Kobayashi, S.} \emph{et~al.}
\newblock \bibinfo{title}{Determining the absolute abundances of natural radioactive elements on the lunar surface by the kaguya gamma-ray spectrometer}.
\newblock \emph{\bibinfo{journal}{Space Sci. Rev.}} \textbf{\bibinfo{volume}{154}}, \bibinfo{pages}{193--218} (\bibinfo{year}{2010}).

\bibitem{Peplowski2011b}
\bibinfo{author}{Peplowski, P.~N.} \emph{et~al.}
\newblock \bibinfo{title}{Radioactive elements on {{Mercury}}'s surface from {{MESSENGER}}: {{Implications}} for the planet's formation and evolution}.
\newblock \emph{\bibinfo{journal}{Science}} \textbf{\bibinfo{volume}{333}}, \bibinfo{pages}{1850--1852} (\bibinfo{year}{2011}).

\bibitem{Peplowski2016c}
\bibinfo{author}{Peplowski, P.~N.}
\newblock \bibinfo{title}{The global elemental composition of 433 {{Eros}}: {{First}} results from the {{NEAR}} gamma-ray spectrometer orbital dataset}.
\newblock \emph{\bibinfo{journal}{Planet. Space Sci.}} \textbf{\bibinfo{volume}{134}}, \bibinfo{pages}{36--51} (\bibinfo{year}{2016}).

\bibitem{Prettyman2017}
\bibinfo{author}{Prettyman, T.~H.} \emph{et~al.}
\newblock \bibinfo{title}{Extensive water ice within {{Ceres}}' aqueously altered regolith: {{Evidence}} from nuclear spectroscopy}.
\newblock \emph{\bibinfo{journal}{Science}} \textbf{\bibinfo{volume}{355}}, \bibinfo{pages}{55--59} (\bibinfo{year}{2017}).

\bibitem{Prettyman2019a}
\bibinfo{author}{Prettyman, T.~H.} \emph{et~al.}
\newblock \bibinfo{title}{Elemental composition and mineralogy of {{Vesta}} and {{Ceres}}: {{Distribution}} and origins of hydrogen-bearing species}.
\newblock \emph{\bibinfo{journal}{Icarus}} \textbf{\bibinfo{volume}{318}}, \bibinfo{pages}{42--55} (\bibinfo{year}{2019}).

\bibitem{Sanada2015a}
\bibinfo{author}{Sanada, Y.} \& \bibinfo{author}{Torii, T.}
\newblock \bibinfo{title}{Aerial radiation monitoring around the {{Fukushima Dai-ichi}} nuclear power plant using an unmanned helicopter}.
\newblock \emph{\bibinfo{journal}{J. Environ. Radioact.}} \textbf{\bibinfo{volume}{139}}, \bibinfo{pages}{294--299} (\bibinfo{year}{2015}).

\bibitem{Connor2016a}
\bibinfo{author}{Connor, D.}, \bibinfo{author}{Martin, P.~G.} \& \bibinfo{author}{Scott, T.~B.}
\newblock \bibinfo{title}{Airborne radiation mapping: Overview and application of current and future aerial systems}.
\newblock \emph{\bibinfo{journal}{Int. J. Remote Sens.}} \textbf{\bibinfo{volume}{37}}, \bibinfo{pages}{5953--5987} (\bibinfo{year}{2016}).

\bibitem{Naumenko2018}
\bibinfo{author}{Naumenko, A.} \emph{et~al.}
\newblock \bibinfo{title}{Autonomous {{NaI}}({{Tl}}) gamma-ray spectrometer for in situ underwater measurements}.
\newblock \emph{\bibinfo{journal}{NIM-A}} \textbf{\bibinfo{volume}{908}}, \bibinfo{pages}{97--109} (\bibinfo{year}{2018}).

\bibitem{Chen2020a}
\bibinfo{author}{Chen, C.~M.}, \bibinfo{author}{Sinclair, L.~E.}, \bibinfo{author}{Fortin, R.}, \bibinfo{author}{Coyle, M.} \& \bibinfo{author}{Samson, C.}
\newblock \bibinfo{title}{In-flight performance of the {{Advanced Radiation Detector}} for {{UAV Operations}} ({{ARDUO}})}.
\newblock \emph{\bibinfo{journal}{Nucl. Instrum. Methods Phys. Res. Sect. Accel. Spectrometers Detect. Assoc. Equip.}} \textbf{\bibinfo{volume}{954}}, \bibinfo{pages}{161609} (\bibinfo{year}{2020}).

\bibitem{Sinclair2021}
\bibinfo{author}{Sinclair, L.~E.} \& \bibinfo{author}{Chen, C.~M.}
\newblock \bibinfo{title}{Inspection of a {{Restricted Access Site Using UAV Perimeter Survey}} with the {{ARDUO Direction-Capable Gamma Spectrometer}}}.
\newblock \emph{\bibinfo{journal}{Pure Appl. Geophys.}} \textbf{\bibinfo{volume}{178}}, \bibinfo{pages}{2779--2788} (\bibinfo{year}{2021}).

\bibitem{Grasty1985a}
\bibinfo{author}{Grasty, R.~L.}, \bibinfo{author}{Glynn, J.~E.} \& \bibinfo{author}{Grant, J.~A.}
\newblock \bibinfo{title}{The analysis of multichannel airborne gamma-ray spectra}.
\newblock \emph{\bibinfo{journal}{Geophysics}} \textbf{\bibinfo{volume}{50}}, \bibinfo{pages}{2611--2620} (\bibinfo{year}{1985}).

\bibitem{Minty1998d}
\bibinfo{author}{Minty, B.~R.}, \bibinfo{author}{McFadden, P.} \& \bibinfo{author}{Kennett, B.~L.}
\newblock \bibinfo{title}{Multichannel processing for airborne gamma-ray spectrometry}.
\newblock \emph{\bibinfo{journal}{Geophysics}} \textbf{\bibinfo{volume}{63}}, \bibinfo{pages}{1971--1985} (\bibinfo{year}{1998}).

\bibitem{Hendriks2001a}
\bibinfo{author}{Hendriks, P.~H.}, \bibinfo{author}{Limburg, J.} \& \bibinfo{author}{De~Meijer, R.~J.}
\newblock \bibinfo{title}{Full-spectrum analysis of natural {$\gamma$}-ray spectra}.
\newblock \emph{\bibinfo{journal}{J. Environ. Radioact.}} \textbf{\bibinfo{volume}{53}}, \bibinfo{pages}{365--380} (\bibinfo{year}{2001}).

\bibitem{Prettyman2011a}
\bibinfo{author}{Prettyman, T.~H.} \emph{et~al.}
\newblock \bibinfo{title}{Dawn's gamma ray and neutron detector}.
\newblock \emph{\bibinfo{journal}{Space Sci. Rev.}} \textbf{\bibinfo{volume}{163}}, \bibinfo{pages}{371--459} (\bibinfo{year}{2011}).

\bibitem{Ohera2024}
\bibinfo{author}{Ohera, M.}, \bibinfo{author}{Gryc, L.}, \bibinfo{author}{Nov{\'a}kov{\'a}, M.}, \bibinfo{author}{{\v C}e{\v s}p{\'i}rov{\'a}, I.} \& \bibinfo{author}{Sas, D.}
\newblock \bibinfo{title}{Application of unmanned aerial vehicles in emergency radiation monitoring}.
\newblock \emph{\bibinfo{journal}{Radiat. Meas.}} \textbf{\bibinfo{volume}{174}}, \bibinfo{pages}{107111} (\bibinfo{year}{2024}).

\bibitem{Sinclair2016a}
\bibinfo{author}{Sinclair, L.~E.} \emph{et~al.}
\newblock \bibinfo{title}{Aerial {{Mobile Radiation Survey Following Detonation}} of a {{Radiological Dispersal Device}}}.
\newblock \emph{\bibinfo{journal}{Health Phys.}} \textbf{\bibinfo{volume}{110}}, \bibinfo{pages}{458--470} (\bibinfo{year}{2016}).

\bibitem{Grasty1991a}
\bibinfo{author}{Grasty, R.~L.}, \bibinfo{author}{Holman, P.~B.} \& \bibinfo{author}{Blanchard, Y.~B.}
\newblock \bibinfo{title}{Transportable calibration pads for ground and airborne gamma-ray spectrometers}.
\newblock \emph{\bibinfo{journal}{Geol. Surv. Can.}} \textbf{\bibinfo{volume}{90}} (\bibinfo{year}{1991}).

\bibitem{Minty1990a}
\bibinfo{author}{Minty, B. R.~S.}, \bibinfo{author}{Morse, M.~P.} \& \bibinfo{author}{Richardson, L.~M.}
\newblock \bibinfo{title}{Portable {{Calibration Sources For Airborne Gamma-ray Spectrometers}}}.
\newblock \emph{\bibinfo{journal}{Explor. Geophys.}} \textbf{\bibinfo{volume}{21}}, \bibinfo{pages}{187--195} (\bibinfo{year}{1990}).

\bibitem{Allyson1998a}
\bibinfo{author}{Allyson, J.~D.} \& \bibinfo{author}{Sanderson, D.~C.}
\newblock \bibinfo{title}{Monte {{Carlo}} simulation of environmental airborne gamma-spectrometry}.
\newblock \emph{\bibinfo{journal}{J. Environ. Radioact.}} \textbf{\bibinfo{volume}{38}}, \bibinfo{pages}{259--282} (\bibinfo{year}{1998}).

\bibitem{Torii2013a}
\bibinfo{author}{Torii, T.}, \bibinfo{author}{Sugita, T.}, \bibinfo{author}{Okada, C.~E.}, \bibinfo{author}{Reed, M.~S.} \& \bibinfo{author}{Blumenthal, D.~J.}
\newblock \bibinfo{title}{Enhanced {{Analysis Methods}} to {{Derive}} the {{Spatial Distribution}} of {{131I Deposition}} on the {{Ground}} by {{Airborne Surveys}} at an {{Early Stage}} after the {{Fukushima Daiichi Nuclear Power Plant Accident}}}.
\newblock \emph{\bibinfo{journal}{Health Phys.}} \textbf{\bibinfo{volume}{105}}, \bibinfo{pages}{192--200} (\bibinfo{year}{2013}).

\bibitem{Kulisek2018a}
\bibinfo{author}{Kulisek, J.~A.} \emph{et~al.}
\newblock \bibinfo{title}{A {{3D}} simulation look-up library for real-time airborne gamma-ray spectroscopy}.
\newblock \emph{\bibinfo{journal}{Nucl. Instrum. Methods Phys. Res. Sect. Accel. Spectrometers Detect. Assoc. Equip.}} \textbf{\bibinfo{volume}{879}}, \bibinfo{pages}{84--91} (\bibinfo{year}{2018}).

\bibitem{Breitenmoser2025a}
\bibinfo{author}{Breitenmoser, D.}, \bibinfo{author}{Stabilini, A.}, \bibinfo{author}{Kasprzak, M.~M.} \& \bibinfo{author}{Mayer, S.}
\newblock \bibinfo{title}{Development and validation of a high-fidelity full-spectrum {{Monte Carlo}} model for the {{Swiss}} airborne gamma-ray spectrometry system}.
\newblock \emph{\bibinfo{journal}{NIM-A}} \textbf{\bibinfo{volume}{1077}}, \bibinfo{pages}{170512} (\bibinfo{year}{2025}).

\bibitem{Breitenmoser2026}
\bibinfo{author}{Breitenmoser, D.}, \bibinfo{author}{Stabilini, A.}, \bibinfo{author}{Kasprzak, M.~M.} \& \bibinfo{author}{Mayer, S.}
\newblock \bibinfo{title}{Full-spectrum modeling of mobile gamma-ray spectrometry systems in scattering media}.
\newblock \emph{\bibinfo{journal}{Phys. Rev. Appl.}} \textbf{\bibinfo{volume}{25}}, \bibinfo{pages}{014044} (\bibinfo{year}{2026}).

\bibitem{Minty1992a}
\bibinfo{author}{Minty, B.~R.}
\newblock \bibinfo{title}{Airborne gamma-ray spectrometric background estimation using full spectrum analysis}.
\newblock \emph{\bibinfo{journal}{Geophysics}} \textbf{\bibinfo{volume}{57}}, \bibinfo{pages}{279--287} (\bibinfo{year}{1992}).

\bibitem{Humphrey2009}
\bibinfo{author}{Humphrey, P.~J.}, \bibinfo{author}{Liu, W.} \& \bibinfo{author}{Buote, D.~A.}
\newblock \bibinfo{title}{{$\chi$}2 {{AND POISSONIAN DATA}}: {{BIASES EVEN IN THE HIGH-COUNT REGIME AND HOW TO AVOID THEM}}}.
\newblock \emph{\bibinfo{journal}{ApJ}} \textbf{\bibinfo{volume}{693}}, \bibinfo{pages}{822} (\bibinfo{year}{2009}).

\bibitem{Yamada2019}
\bibinfo{author}{Yamada, S.} \emph{et~al.}
\newblock \bibinfo{title}{Poisson vs. {{Gaussian}} statistics for sparse {{X-ray}} data: {{Application}} to the soft {{X-ray}} spectrometer}.
\newblock \emph{\bibinfo{journal}{Publ Astron Soc Jpn Nihon Tenmon Gakkai}} \textbf{\bibinfo{volume}{71}}, \bibinfo{pages}{75} (\bibinfo{year}{2019}).

\bibitem{Veitch2015}
\bibinfo{author}{Veitch, J.}
\newblock \bibinfo{title}{Parameter estimation for compact binaries with ground-based gravitational-wave observations using the {{LALInference}} software library}.
\newblock \emph{\bibinfo{journal}{Phys. Rev. D}} \textbf{\bibinfo{volume}{91}} (\bibinfo{year}{2015}).

\bibitem{Ashton2019}
\bibinfo{author}{Ashton, G.} \emph{et~al.}
\newblock \bibinfo{title}{Bilby: {{A User-friendly Bayesian Inference Library}} for {{Gravitational-wave Astronomy}}}.
\newblock \emph{\bibinfo{journal}{ApJS}} \textbf{\bibinfo{volume}{241}}, \bibinfo{pages}{27} (\bibinfo{year}{2019}).

\bibitem{Smith2020}
\bibinfo{author}{Smith, R. J.~E.}, \bibinfo{author}{Ashton, G.}, \bibinfo{author}{Vajpeyi, A.} \& \bibinfo{author}{Talbot, C.}
\newblock \bibinfo{title}{Massively parallel {{Bayesian}} inference for transient gravitational-wave astronomy}.
\newblock \emph{\bibinfo{journal}{MNRAS}} \textbf{\bibinfo{volume}{498}}, \bibinfo{pages}{4492--4502} (\bibinfo{year}{2020}).

\bibitem{MacDonald2017}
\bibinfo{author}{MacDonald, R.~J.} \& \bibinfo{author}{Madhusudhan, N.}
\newblock \bibinfo{title}{{{HD}} 209458b in new light: Evidence of nitrogen chemistry, patchy clouds and sub-solar water}.
\newblock \emph{\bibinfo{journal}{MNRAS}} \textbf{\bibinfo{volume}{469}}, \bibinfo{pages}{1979--1996} (\bibinfo{year}{2017}).

\bibitem{Pinhas2018}
\bibinfo{author}{Pinhas, A.}, \bibinfo{author}{Rackham, B.~V.}, \bibinfo{author}{Madhusudhan, N.} \& \bibinfo{author}{Apai, D.}
\newblock \bibinfo{title}{Retrieval of planetary and stellar properties in transmission spectroscopy with {{Aura}}}.
\newblock \emph{\bibinfo{journal}{MNRAS}} \textbf{\bibinfo{volume}{480}}, \bibinfo{pages}{5314--5331} (\bibinfo{year}{2018}).

\bibitem{Trotta2008a}
\bibinfo{author}{Trotta, R.}
\newblock \bibinfo{title}{Bayes in the sky: {{Bayesian}} inference and model selection in cosmology}.
\newblock \emph{\bibinfo{journal}{Contemp. Phys.}} \textbf{\bibinfo{volume}{49}}, \bibinfo{pages}{71--104} (\bibinfo{year}{2008}).

\bibitem{VonToussaint2011a}
\bibinfo{author}{{von Toussaint}, U.}
\newblock \bibinfo{title}{Bayesian inference in physics}.
\newblock \emph{\bibinfo{journal}{Rev. Mod. Phys.}} \textbf{\bibinfo{volume}{83}}, \bibinfo{pages}{943--999} (\bibinfo{year}{2011}).

\bibitem{Romano2015}
\bibinfo{author}{Romano, P.~K.} \emph{et~al.}
\newblock \bibinfo{title}{{{OpenMC}}: {{A}} state-of-the-art {{Monte Carlo}} code for research and development}.
\newblock \emph{\bibinfo{journal}{Ann. Nucl. Energy}} \textbf{\bibinfo{volume}{82}}, \bibinfo{pages}{90--97} (\bibinfo{year}{2015}).

\bibitem{Allison2016}
\bibinfo{author}{Allison, J.} \emph{et~al.}
\newblock \bibinfo{title}{Recent developments in {{Geant4}}}.
\newblock \emph{\bibinfo{journal}{Nucl. Instrum. Methods Phys. Res. A}} \textbf{\bibinfo{volume}{835}}, \bibinfo{pages}{186--225} (\bibinfo{year}{2016}).

\bibitem{Goorley2016}
\bibinfo{author}{Goorley, T.} \emph{et~al.}
\newblock \bibinfo{title}{Features of {{MCNP6}}}.
\newblock \emph{\bibinfo{journal}{Ann. Nucl. Energy}} \textbf{\bibinfo{volume}{87}}, \bibinfo{pages}{772--783} (\bibinfo{year}{2016}).

\bibitem{Ahdida2022a}
\bibinfo{author}{Ahdida, C.} \emph{et~al.}
\newblock \bibinfo{title}{New {{Capabilities}} of the {{FLUKA Multi-Purpose Code}}}.
\newblock \emph{\bibinfo{journal}{Front. Phys.}} \textbf{\bibinfo{volume}{9}}, \bibinfo{pages}{788253} (\bibinfo{year}{2022}).

\bibitem{Sato2024}
\bibinfo{author}{Sato, T.} \emph{et~al.}
\newblock \bibinfo{title}{Recent improvements of the particle and heavy ion transport code system -- {{PHITS}} version 3.33}.
\newblock \emph{\bibinfo{journal}{J. Nucl. Sci. Technol.}} \textbf{\bibinfo{volume}{61}}, \bibinfo{pages}{127--135} (\bibinfo{year}{2024}).

\bibitem{Marelli2014a}
\bibinfo{author}{Marelli, S.} \& \bibinfo{author}{Sudret, B.}
\newblock \bibinfo{title}{{{UQLab}}: {{A Framework}} for {{Uncertainty Quantification}} in {{Matlab}}}.
\newblock \emph{\bibinfo{journal}{ICVRAM}} \bibinfo{pages}{2554--2563} (\bibinfo{year}{2014}).

\bibitem{Carpenter2017}
\bibinfo{author}{Carpenter, B.} \emph{et~al.}
\newblock \bibinfo{title}{Stan: {{A Probabilistic Programming Language}}}.
\newblock \emph{\bibinfo{journal}{J. Stat. Softw.}} \textbf{\bibinfo{volume}{76}}, \bibinfo{pages}{1--32} (\bibinfo{year}{2017}).

\bibitem{Buchner2021c}
\bibinfo{author}{Buchner, J.}
\newblock \bibinfo{title}{{{UltraNest}} - a robust, general purpose {{Bayesian}} inference engine}.
\newblock \emph{\bibinfo{journal}{J. Open Source Softw.}} \textbf{\bibinfo{volume}{6}}, \bibinfo{pages}{3001} (\bibinfo{year}{2021}).

\bibitem{Abril-Pla2023}
\bibinfo{author}{{Abril-Pla}, O.} \emph{et~al.}
\newblock \bibinfo{title}{{{PyMC}}: A modern, and comprehensive probabilistic programming framework in {{Python}}}.
\newblock \emph{\bibinfo{journal}{PeerJ Comput. Sci.}} \textbf{\bibinfo{volume}{9}}, \bibinfo{pages}{e1516} (\bibinfo{year}{2023}).

\bibitem{Gelman2013}
\bibinfo{author}{Gelman, A.} \emph{et~al.}
\newblock \emph{\bibinfo{title}{Bayesian {{Data Analysis}}}} \bibinfo{edition}{3rd} edn (\bibinfo{publisher}{{Chapman and Hall/CRC}}, \bibinfo{address}{New York, USA}, \bibinfo{year}{2013}).

\bibitem{DAgostini2003a}
\bibinfo{author}{D'Agostini, G.}
\newblock \bibinfo{title}{Bayesian inference in processing experimental data: Principles and basic applications}.
\newblock \emph{\bibinfo{journal}{Rep. Prog. Phys.}} \textbf{\bibinfo{volume}{66}}, \bibinfo{pages}{1383} (\bibinfo{year}{2003}).

\bibitem{Breitenmoser2025j}
\bibinfo{author}{Breitenmoser, D.}, \bibinfo{author}{Lopez, R.}, \bibinfo{author}{Clarke, S.~D.} \& \bibinfo{author}{Pozzi, S.~A.}
\newblock \bibinfo{title}{Identifying neutron sources using recoil and time-of-flight spectroscopy}.
\newblock \emph{\bibinfo{journal}{Phys. Rev. Appl.}} \textbf{\bibinfo{volume}{25}}, \bibinfo{pages}{064013} (\bibinfo{year}{2026}).

\bibitem{Praszalowicz2011}
\bibinfo{author}{Praszalowicz, M.}
\newblock \bibinfo{title}{Negative {{Binomial Distribution}} and the multiplicity moments at the {{LHC}}}.
\newblock \emph{\bibinfo{journal}{Phys. Lett. B}} \textbf{\bibinfo{volume}{704}}, \bibinfo{pages}{566--569} (\bibinfo{year}{2011}).

\bibitem{Tezlaf2023}
\bibinfo{author}{Tezlaf, S.~V.}
\newblock \bibinfo{title}{Significance of the negative binomial distribution in multiplicity phenomena}.
\newblock \emph{\bibinfo{journal}{Phys. Scr.}} \textbf{\bibinfo{volume}{98}}, \bibinfo{pages}{115310} (\bibinfo{year}{2023}).

\bibitem{Perez2021}
\bibinfo{author}{Perez, L.~A.}, \bibinfo{author}{Malhotra, S.}, \bibinfo{author}{Rhoads, J.~E.} \& \bibinfo{author}{Tilvi, V.}
\newblock \bibinfo{title}{Void {{Probability Function}} of {{Simulated Surveys}} of {{High-redshift Ly$\alpha$ Emitters}}}.
\newblock \emph{\bibinfo{journal}{ApJ}} \textbf{\bibinfo{volume}{906}}, \bibinfo{pages}{58} (\bibinfo{year}{2021}).

\bibitem{Fry2013}
\bibinfo{author}{Fry, J.~N.} \& \bibinfo{author}{Colombi, S.}
\newblock \bibinfo{title}{Void statistics and hierarchical scaling in the halo model}.
\newblock \emph{\bibinfo{journal}{MNRAS}} \textbf{\bibinfo{volume}{433}}, \bibinfo{pages}{581--590} (\bibinfo{year}{2013}).

\bibitem{Hurtado-Gil2017}
\bibinfo{author}{{Hurtado-Gil}, L.} \emph{et~al.}
\newblock \bibinfo{title}{The best fit for the observed galaxy counts-in-cell distribution function}.
\newblock \emph{\bibinfo{journal}{A\&A}} \textbf{\bibinfo{volume}{601}}, \bibinfo{pages}{A40} (\bibinfo{year}{2017}).

\bibitem{Hameeda2021}
\bibinfo{author}{Hameeda, M.}, \bibinfo{author}{Plastino, A.} \& \bibinfo{author}{Rocca, M.~C.}
\newblock \bibinfo{title}{Generalized {{Poisson}} distributions for systems with two-particle interactions}.
\newblock \emph{\bibinfo{journal}{IOPSciNotes}} \textbf{\bibinfo{volume}{2}}, \bibinfo{pages}{015003} (\bibinfo{year}{2021}).

\bibitem{Metropolis1953}
\bibinfo{author}{Metropolis, N.}, \bibinfo{author}{Rosenbluth, A.~W.}, \bibinfo{author}{Rosenbluth, M.~N.}, \bibinfo{author}{Teller, A.~H.} \& \bibinfo{author}{Teller, E.}
\newblock \bibinfo{title}{Equation of {{State Calculations}} by {{Fast Computing Machines}}}.
\newblock \emph{\bibinfo{journal}{J. Chem. Phys.}} \textbf{\bibinfo{volume}{21}}, \bibinfo{pages}{1087--1092} (\bibinfo{year}{1953}).

\bibitem{Hastings1970}
\bibinfo{author}{Hastings, W.~K.}
\newblock \bibinfo{title}{Monte {{Carlo}} sampling methods using {{Markov}} chains and their applications}.
\newblock \emph{\bibinfo{journal}{Biometrika}} \textbf{\bibinfo{volume}{57}}, \bibinfo{pages}{97--109} (\bibinfo{year}{1970}).

\bibitem{Foreman-Mackey2013a}
\bibinfo{author}{{Foreman-Mackey}, D.}, \bibinfo{author}{Hogg, D.~W.}, \bibinfo{author}{Lang, D.} \& \bibinfo{author}{Goodman, J.}
\newblock \bibinfo{title}{Emcee : {{The MCMC Hammer}}}.
\newblock \emph{\bibinfo{journal}{Publ. Astron. Soc. Pac.}} \textbf{\bibinfo{volume}{125}}, \bibinfo{pages}{306--312} (\bibinfo{year}{2013}).

\bibitem{Goodman2010}
\bibinfo{author}{Goodman, J.} \& \bibinfo{author}{Weare, J.}
\newblock \bibinfo{title}{Ensemble samplers with affine invariance}.
\newblock \emph{\bibinfo{journal}{Commun. Appl. Math. Comput. Sci.}} \textbf{\bibinfo{volume}{5}}, \bibinfo{pages}{65--80} (\bibinfo{year}{2010}).

\bibitem{Perrakis2014}
\bibinfo{author}{Perrakis, K.}, \bibinfo{author}{Ntzoufras, I.} \& \bibinfo{author}{Tsionas, E.~G.}
\newblock \bibinfo{title}{On the use of marginal posteriors in marginal likelihood estimation via importance sampling}.
\newblock \emph{\bibinfo{journal}{Comput. Stat. Data Anal.}} \textbf{\bibinfo{volume}{77}}, \bibinfo{pages}{54--69} (\bibinfo{year}{2014}).

\bibitem{Metodiev2024}
\bibinfo{author}{Metodiev, M.} \emph{et~al.}
\newblock \bibinfo{title}{Easily {{Computed Marginal Likelihoods}} from {{Posterior Simulation Using}} the {{THAMES Estimator}}}.
\newblock \emph{\bibinfo{journal}{Bayesian Anal.}} \textbf{\bibinfo{volume}{-1}}, \bibinfo{pages}{1--28} (\bibinfo{year}{2024}).

\bibitem{Llorente2023}
\bibinfo{author}{Llorente, F.}, \bibinfo{author}{Martino, L.}, \bibinfo{author}{Delgado, D.} \& \bibinfo{author}{{L{\'o}pez-Santiago}, J.}
\newblock \bibinfo{title}{Marginal {{Likelihood Computation}} for {{Model Selection}} and {{Hypothesis Testing}}: {{An Extensive Review}}}.
\newblock \emph{\bibinfo{journal}{SIAM Rev.}} \textbf{\bibinfo{volume}{65}}, \bibinfo{pages}{3--58} (\bibinfo{year}{2023}).

\bibitem{Skilling2006a}
\bibinfo{author}{Skilling, J.}
\newblock \bibinfo{title}{Nested sampling for general {{Bayesian}} computation}.
\newblock \emph{\bibinfo{journal}{Bayesian Anal.}} \textbf{\bibinfo{volume}{1}}, \bibinfo{pages}{833--859} (\bibinfo{year}{2006}).

\bibitem{Feroz2009a}
\bibinfo{author}{Feroz, F.}, \bibinfo{author}{Hobson, M.~P.} \& \bibinfo{author}{Bridges, M.}
\newblock \bibinfo{title}{{{MultiNest}}: {{An}} efficient and robust {{Bayesian}} inference tool for cosmology and particle physics}.
\newblock \emph{\bibinfo{journal}{Mon. Not. R. Astron. Soc.}} \textbf{\bibinfo{volume}{398}}, \bibinfo{pages}{1601--1614} (\bibinfo{year}{2009}).

\bibitem{Speagle2020a}
\bibinfo{author}{Speagle, J.~S.}
\newblock \bibinfo{title}{{{DYNESTY}}: A dynamic nested sampling package for estimating {{Bayesian}} posteriors and evidences}.
\newblock \emph{\bibinfo{journal}{Mon. Not. R. Astron. Soc.}} \textbf{\bibinfo{volume}{493}}, \bibinfo{pages}{3132--3158} (\bibinfo{year}{2020}).

\bibitem{Ashton2022a}
\bibinfo{author}{Ashton, G.} \emph{et~al.}
\newblock \bibinfo{title}{Nested sampling for physical scientists}.
\newblock \emph{\bibinfo{journal}{Nat. Rev. Methods Primer 2022 21}} \textbf{\bibinfo{volume}{2}}, \bibinfo{pages}{1--22} (\bibinfo{year}{2022}).

\bibitem{VasilEv2014a}
\bibinfo{author}{Vasil'Ev, A.~N.} \& \bibinfo{author}{Gektin, A.~V.}
\newblock \bibinfo{title}{Multiscale approach to estimation of scintillation characteristics}.
\newblock \emph{\bibinfo{journal}{IEEE Trans. Nucl. Sci.}} \textbf{\bibinfo{volume}{61}}, \bibinfo{pages}{235--245} (\bibinfo{year}{2014}).

\bibitem{Breitenmoser2023c}
\bibinfo{author}{Breitenmoser, D.}, \bibinfo{author}{Cerutti, F.}, \bibinfo{author}{Butterweck, G.}, \bibinfo{author}{Kasprzak, M.~M.} \& \bibinfo{author}{Mayer, S.}
\newblock \bibinfo{title}{Emulator-based {{Bayesian}} inference on non-proportional scintillation models by compton-edge probing}.
\newblock \emph{\bibinfo{journal}{Nat. Commun.}} \textbf{\bibinfo{volume}{14}}, \bibinfo{pages}{7790} (\bibinfo{year}{2023}).

\bibitem{Breitenmoser2022}
\bibinfo{author}{Breitenmoser, D.}, \bibinfo{author}{Butterweck, G.}, \bibinfo{author}{Kasprzak, M.~M.}, \bibinfo{author}{Yukihara, E.~G.} \& \bibinfo{author}{Mayer, S.}
\newblock \bibinfo{title}{Experimental and {{Simulated Spectral Gamma-Ray Response}} of a {{NaI}}({{Tl}}) {{Scintillation Detector}} used in {{Airborne Gamma-Ray Spectrometry}}}.
\newblock \emph{\bibinfo{journal}{Adv. Geosci.}} \textbf{\bibinfo{volume}{57}}, \bibinfo{pages}{89--107} (\bibinfo{year}{2022}).

\bibitem{Butterweck2023}
\bibinfo{author}{Butterweck, G.} \emph{et~al.}
\newblock \bibinfo{title}{Aeroradiometric measurements in the framework of the {{Swiss Exercise ARM22}}}.
\newblock \bibinfo{type}{Tech. Rep.}, \bibinfo{institution}{Paul Scherrer Institut (PSI)}, \bibinfo{address}{Villigen PSI, Switzerland} (\bibinfo{year}{2023}).

\bibitem{Butterweck2021a}
\bibinfo{author}{Butterweck, G.} \emph{et~al.}
\newblock \bibinfo{title}{Aeroradiometric measurements in the framework of the {{Swiss}} exercise {{ARM20}}}.
\newblock \bibinfo{type}{Tech. Rep.}, \bibinfo{institution}{Paul Scherrer Institut (PSI)}, \bibinfo{address}{Villigen PSI, Switzerland} (\bibinfo{year}{2021}).

\bibitem{Lyons2012}
\bibinfo{author}{Lyons, C.} \& \bibinfo{author}{Colton, D.}
\newblock \bibinfo{title}{Aerial {{Measuring System}} in {{Japan}}}.
\newblock \emph{\bibinfo{journal}{Health Phys.}} \textbf{\bibinfo{volume}{102}}, \bibinfo{pages}{509} (\bibinfo{year}{2012}).

\bibitem{Winkelmann2004a}
\bibinfo{author}{Winkelmann, I.}, \bibinfo{author}{Strobl, C.} \& \bibinfo{author}{Thomas, M.}
\newblock \bibinfo{title}{Aerial measurements of artificial radionuclides in {{Germany}} in case of a nuclear accident}.
\newblock \emph{\bibinfo{journal}{J. Environ. Radioact.}} \textbf{\bibinfo{volume}{72}}, \bibinfo{pages}{225--231} (\bibinfo{year}{2004}).

\bibitem{Sellke2001}
\bibinfo{author}{Sellke, T.}, \bibinfo{author}{Bayarri, M.~J.} \& \bibinfo{author}{Berger, J.~O.}
\newblock \bibinfo{title}{Calibration of {$\rho$} {{Values}} for {{Testing Precise Null Hypotheses}}}.
\newblock \emph{\bibinfo{journal}{Am. Stat.}} \textbf{\bibinfo{volume}{55}}, \bibinfo{pages}{62--71} (\bibinfo{year}{2001}).

\bibitem{Jeffreys1948}
\bibinfo{author}{Jeffreys, H.}
\newblock \emph{\bibinfo{title}{Theory {{Of Probability}}}} \bibinfo{edition}{2} edn (\bibinfo{publisher}{Oxford University Press}, \bibinfo{year}{1948}).

\bibitem{Jandel2004a}
\bibinfo{author}{Jandel, M.} \emph{et~al.}
\newblock \bibinfo{title}{Decomposition of continuum {$\gamma$}-ray spectra using synthesized response matrix}.
\newblock \emph{\bibinfo{journal}{Nucl. Instrum. Methods Phys. Res. Sect. Accel. Spectrometers Detect. Assoc. Equip.}} \textbf{\bibinfo{volume}{516}}, \bibinfo{pages}{172--183} (\bibinfo{year}{2004}).

\bibitem{Baldoncini2018b}
\bibinfo{author}{Baldoncini, M.} \emph{et~al.}
\newblock \bibinfo{title}{Investigating the potentialities of {{Monte Carlo}} simulation for assessing soil water content via proximal gamma-ray spectroscopy}.
\newblock \emph{\bibinfo{journal}{J. Environ. Radioact.}} \textbf{\bibinfo{volume}{192}}, \bibinfo{pages}{105--116} (\bibinfo{year}{2018}).

\bibitem{Kastlander2005a}
\bibinfo{author}{Kastlander, J.} \& \bibinfo{author}{Bargholtz, C.}
\newblock \bibinfo{title}{Efficient in situ method to determine radionuclide concentration in soil}.
\newblock \emph{\bibinfo{journal}{Nucl. Instrum. Methods Phys. Res. Sect. Accel. Spectrometers Detect. Assoc. Equip.}} \textbf{\bibinfo{volume}{547}}, \bibinfo{pages}{400--410} (\bibinfo{year}{2005}).

\bibitem{Namito2012a}
\bibinfo{author}{Namito, Y.} \emph{et~al.}
\newblock \bibinfo{title}{Transformation of a system consisting of plane isotropic source and unit sphere detector into a system consisting of point isotropic source and plane detector in {{Monte Carlo}} radiation transport calculation}.
\newblock \emph{\bibinfo{journal}{J. Nucl. Sci. Technol.}} \textbf{\bibinfo{volume}{49}}, \bibinfo{pages}{167--172} (\bibinfo{year}{2012}).

\bibitem{McConn2011a}
\bibinfo{author}{McConn, R.~J.}, \bibinfo{author}{Gesh, C.~J.}, \bibinfo{author}{Pagh, R.~T.} \& \bibinfo{author}{Rucker, R.~A.}
\newblock \bibinfo{title}{Compendium of {{Material Composition Data}} for {{Radiation Transport Modeling}}}.
\newblock \bibinfo{type}{Tech. Rep.}, \bibinfo{institution}{Pacific Northwest National Laboratory}, \bibinfo{address}{Richland} (\bibinfo{year}{2011}).

\bibitem{Hunnefeld2022a}
\bibinfo{author}{H{\"u}nnefeld, M.} \emph{et~al.}
\newblock \bibinfo{title}{Combining {{Maximum-Likelihood}} with {{Deep Learning}} for {{Event Reconstruction}} in {{IceCube}}}.
\newblock \emph{\bibinfo{journal}{Proc. Sci.}} \textbf{\bibinfo{volume}{395}}, \bibinfo{pages}{1065} (\bibinfo{year}{2022}).

\bibitem{Salinas2020a}
\bibinfo{author}{Salinas, D.}, \bibinfo{author}{Flunkert, V.}, \bibinfo{author}{Gasthaus, J.} \& \bibinfo{author}{Januschowski, T.}
\newblock \bibinfo{title}{{{DeepAR}}: {{Probabilistic}} forecasting with autoregressive recurrent networks}.
\newblock \emph{\bibinfo{journal}{Int. J. Forecast.}} \textbf{\bibinfo{volume}{36}}, \bibinfo{pages}{1181--1191} (\bibinfo{year}{2020}).

\bibitem{Lloyd-Smith2007a}
\bibinfo{author}{{Lloyd-Smith}, J.~O.}
\newblock \bibinfo{title}{Maximum {{Likelihood Estimation}} of the {{Negative Binomial Dispersion Parameter}} for {{Highly Overdispersed Data}}, with {{Applications}} to {{Infectious Diseases}}}.
\newblock \emph{\bibinfo{journal}{PLOS ONE}} \textbf{\bibinfo{volume}{2}}, \bibinfo{pages}{e180} (\bibinfo{year}{2007}).

\bibitem{Paradis2020a}
\bibinfo{author}{Paradis, H.} \emph{et~al.}
\newblock \bibinfo{title}{Spectral unmixing applied to fast identification of {$\gamma$}-emitting radionuclides using {{NaI}}({{Tl}}) detectors}.
\newblock \emph{\bibinfo{journal}{Appl. Radiat. Isot.}} \textbf{\bibinfo{volume}{158}}, \bibinfo{pages}{109068} (\bibinfo{year}{2020}).

\bibitem{Andre2021a}
\bibinfo{author}{Andr{\'e}, R.}, \bibinfo{author}{Bobin, C.}, \bibinfo{author}{Bobin, J.}, \bibinfo{author}{Xu, J.} \& \bibinfo{author}{{de Vismes Ott}, A.}
\newblock \bibinfo{title}{Metrological approach of {$\gamma$}-emitting radionuclides identification at low statistics: {{Application}} of sparse spectral unmixing to scintillation detectors}.
\newblock \emph{\bibinfo{journal}{Metrologia}} \textbf{\bibinfo{volume}{58}}, \bibinfo{pages}{15011--15025} (\bibinfo{year}{2021}).

\bibitem{Brooks1998a}
\bibinfo{author}{Brooks, S.~P.} \& \bibinfo{author}{Gelman, A.}
\newblock \bibinfo{title}{General {{Methods}} for {{Monitoring Convergence}} of {{Iterative Simulations}}}.
\newblock \emph{\bibinfo{journal}{J. Comput. Graph. Stat.}} \textbf{\bibinfo{volume}{7}}, \bibinfo{pages}{434--455} (\bibinfo{year}{1998}).

\bibitem{Breitenmoser2025k}
\bibinfo{author}{Breitenmoser, D.}, \bibinfo{author}{Stabilini, A.}, \bibinfo{author}{Kasprzak, M.~M.} \& \bibinfo{author}{Mayer, S.}
\newblock \bibinfo{title}{Quantitative mobile gamma-ray spectrometry through {{Bayesian}} inference ({{Dataset}})}. 
\newblock \emph{\bibinfo{journal}{Zenodo}} (\bibinfo{year}{2025}).
\newblock \doi{10.5281/zenodo.18004440}.

\bibitem{Vlachoudis2009a}
\bibinfo{author}{Vlachoudis, V.}
\newblock \bibinfo{title}{Flair: {{A}} powerful but user friendly graphical interface for {{FLUKA}}}.
\newblock \emph{\bibinfo{journal}{Int. Conf. Math. Comput. Methods React. Phys. MC 2009}}  (\bibinfo{year}{2009}).

\bibitem{Breitenmoser2023}
\bibinfo{author}{Breitenmoser, D.}, \bibinfo{author}{Cerutti, F.}, \bibinfo{author}{Butterweck, G.}, \bibinfo{author}{Kasprzak, M.~M.} \& \bibinfo{author}{Mayer, S.}
\newblock \bibinfo{title}{{{FLUKA}} user routines for non-proportional scintillation simulations}. 
\newblock \emph{\bibinfo{journal}{ETH Research Collection}} (\bibinfo{year}{2023}).
\newblock \doi{10.3929/ETHZ-B-000595727}.

\end{thebibliography}
\end{document}